\documentclass[12pt,preprint]{aastex}

\newcommand{\be}{\begin{equation}}
\newcommand{\ee}{\end{equation}}
\def\bea{\begin{eqnarray}}
\def\eea{\end{eqnarray}}
\newcommand\eps{\varepsilon_{\star}}  
\newcommand{\cs}{c_{\star}}  
\newcommand{\epg}{\varepsilon_{g}}  
\newcommand{\msu}{M_{\odot}}  
\newcommand{\La}{{$\Lambda$CDM}}
\newcommand{\M}{{CHDM}}
\newcommand{\gr}{\kern 2pt\hbox{}^\circ{\kern -2pt K}} 
\newcommand{\oml}{\Omega_{\Lambda}}
\newcommand{\omm}{\Omega_{m}}
\newcommand{\brr}{\begin{array}}
\newcommand{\err}{\end{array}}
\newcommand{\ltsima}{$\; \buildrel < \over \sim \;$}
\newcommand{\simlt}{\lower.5ex\hbox{\ltsima}}
\newcommand{\gtsima}{$\; \buildrel > \over \sim \;$}
\newcommand{\simgt}{\lower.5ex\hbox{\gtsima}}
\shorttitle{SPH simulations of cooling clusters}
\shortauthors{Valdarnini}
\begin{document} 
\title{Numerical convergence of physical variables in  \\
  hydrodynamical simulations of cooling clusters}
\author{R. Valdarnini\altaffilmark{1} }
\affil{SISSA Via Beirut 2-4 34014, Trieste, Italy}
\email{valda@sissa.it}
\begin{abstract}
Results from hydrodynamical SPH simulations of galaxy clusters are used
to investigate the dependence of the final cluster X-ray properties 
on the numerical resolution and the assumed models for the physical gas
processes. Two different spatially flat cosmological models have been 
considered: a low-density cold dark matter universe with a vacuum 
energy density $\oml=0.7$ ($\Lambda$CDM) and a cold+hot dark matter model
(CHDM). For each of these models two different clusters have been extracted
from a cosmological $N-$body simulation.  
A  series of hydrodynamical simulations has then been performed for each of
them using a TREESPH code.
These simulations first include radiative cooling and then also 
conversion of cold gas particles into stars; because of supernova explosions 
these particles can release energy in the form of thermal energy to the
surrounding intracluster gas.
For a specific treatment for the thermal state of the gas, simulation runs 
have been
performed with different numerical resolutions. This is in order to disentangle 
in the final results for the cluster profiles, the effects  of the resolution
from those due to the assumed model for the gas thermal evolution.
The numerical resolution of the simulation is controlled by the 
 number of gas particles $N_g$ and the chosen value 
for the gas gravitational softening parameter $\epg$. The latter is proportional to
the minimum SPH smoothing length and therefore sets a
maximum spatial resolution for the simulations.
For the cooling runs, final X-ray luminosities have been found to be diverging
according to $ L_X \propto 1/\epg^{\sim 5}$. The gas density profiles 
are also diverging at the cluster center. 
This is in agreement with previous findings. When cold gas particles are allowed
to convert into stars, the divergences are removed. The final gas profiles show
a well defined core radius and the temperature profiles are nearly flat. 
For the most massive test cluster in the $\Lambda$CDM model, these simulations
show a prominent cooling flow in the cluster core.
This cluster was analyzed in detail, running simulations with 
different star formation methods and increasing numerical resolution.
A comparison between different runs shows that the results of simulations,
based on star formation methods in which gas conversion into stars is
controlled by an efficiency parameter $c_{\star}$, are sensitive to the
numerical resolution of the simulation.
In this respect star formation methods based instead on a local density
threshold, as in Navarro and White (1993), are shown to give more stable 
results. Final X-ray luminosities are found to be numerically stable,
with uncertainties of a factor $\sim 2$.
These simulations are also in good agreement with observational data when 
the final results are compared with the observed star formation rate and the 
luminosity-temperature relation from cooling flow clusters. Therefore I find 
that hydrodynamical simulations of cooling clusters can be used to give
reliably predictions of the 
 cluster X-ray properties. For a given numerical resolution,
 the conversion of cool gas particles into stars as in Navarro and White should 
be preferred.
\end{abstract}

\section{Introduction}

Galaxy clusters are the largest virialized structures known in the universe.
According to the hierarchical scenario their evolution rate is a
strong function of the background cosmology \citep{peb89,lil92,
via99,ouk92,ekea98,sad98,bah97}, 
 thus making galaxy clusters natural tools for constraining  the
 cosmological models.
Galaxy clusters are also powerful X-ray emitters.
X-ray observations have shown that most of the baryons in galaxy 
clusters are in the form of hot ( $T \simeq 10^7 \gr $ ) ionized 
X-ray emitting gas \citep{for82,sar86}.
The bulk of the emission is via thermal bremsstrahlung; 
its dependence on the square of the gas density allows one to select
cluster samples without the contamination effects which may arise 
in the optical band. For this reason, galaxy clusters have been the 
subject of extensive observational programs in the X-ray band 
\citep{hen92,hen97,ebe97,ros98}.
Observations of the cluster X-ray emission and temperature can be
used to reconstruct the radial gas density and temperature profile,
 assuming spherical symmetry.
The gas density profile has a radial fall-off with slope $ \simeq {-2}$ and 
 a constant density core in the inner regions, with typical size  
$r_g \simeq 50-200 h^{-1} Kpc$ \citep{sar86}; the gas temperature profile 
 is instead nearly isothermal within the virial radius.
The density of the gas can be used to recover the dark matter profile
and the total cluster mass. Assuming virial equilibrium these cluster 
properties are
connected to the primeval power spectrum and the assumed cosmological model, 
thus providing important clues for testing cosmologies
\citep{mak98,nfw97}.

Other cosmological information can be obtained from the statistical 
properties of the ensemble of X-ray clusters.
X-ray observations of cluster number counts, the X-ray temperature function
\citep{hen91,edg90,hen97} and the X-ray luminosity function 
 \citep{ros98,ebe98} are powerful probes for constraining the values of
the cosmological parameters $\Omega_0$ and $\sigma_8$ 
 \citep{hen91,whi93,bah98,eke96,kit98}.
An analytical framework for connecting the X-ray temperature and the luminosity 
function to theoretical 
models can be obtained, within the Press \& Schecter (1974) approach,
by assuming hydrostatic equilibrium for the gas distribution 
and neglecting radiative cooling.
In accordance with these assumptions, analytical methods can then be used 
to derive  predictions for the evolution of the X-ray 
luminosity function and its correlation with the cluster X-ray temperature
 \citep{kai86,kit96,mat98,via99}.

Given the wealth of information which can be obtained from observations
of X-ray clusters, a lot of efforts have been devoted to obtaining directly 
the gas and temperature distributions, using numerical simulations
for investigating the evolution of galaxy clusters.
Collisionless N-body simulations have been used to study substructure 
formation \citep{wes88,cro96,bx97,jin95}, 
density profiles \citep{cro94,hus99,mor98,jin00},
 and statistical properties in a cosmological volume \citep{eke96,lac94}.
 In these simulations the dark matter distribution is a good
tracer of that of the gas provided that the cluster dynamical state 
is not far from equilibrium, with a small fraction of substructures 
within the virial radius. Furthermore, the dark matter density profile shows no 
evidence for a core radius \citep{mor98} whatever numerical resolution 
is achieved.
Numerical simulations have then been extended to include the hydrodynamics
of the gaseous component. The numerical methods used are either 
Eulerian  \citep{cen94,ann96,bry98,kan94,bry94,cen97,cen95},
with a fixed or adaptative grid, or Lagrangian 
\citep{evr88,evr90,tho92,nfw95,ekeb98,kat93,yos98,val99}.
The Lagrangian schemes are based on the 
smoothed  particle hydrodynamic technique \citep{gin77,mon92}.
In these simulations the required dynamical range can be quite demanding.
A large simulation box is needed in order to obtain a meaningful 
statistical sample of galaxy clusters; at the same time the minimum
spatial resolution must be at least close to the cluster core radii,
where the bulk of the X-ray emission originates.
For Lagrangian methods a partial solution is the multimass technique 
\citep{kat93}, where the particles increase their masses according to their
distance from the cluster center. A single cluster can then be 
simulated with a comparatively high numerical resolution, with external shells
of matter surrounding the cluster and representing the large-scale 
gravitational field.
In these simulations the gas component is treated as a single adiabatic fluid,
without taking into account the effects of radiative cooling, and the physical
processes of merging, substructure formation, shocks
and compressional heating of the gas can be modelled in this way. A comparison 
between different 
numerical simulations shows that they are successful in reproducing 
the gross features of the cluster properties \citep{fre99}.

The inclusion of radiative cooling for the gas is important on scales where 
the gas cooling time is shorter than the Hubble time. For galaxy clusters 
the spatial extent of this region is between $50$ and $200~Kpc$ from the
cluster center. A gas cooling radius $r_{cool}$ can then be defined where
the two time scales are equal.
There is of observational evidence that numerical simulations must
include radiative cooling, together with star formation and energy 
feedback, in order to model adequately the relevant physical processes of the 
gas during its evolution.
In the inner regions where the radiative time scales are short, a cooling flow
develops (Fabian 1994), with $r_{cool} \simeq 50 h^{-1}Kpc$.
More than $ 50 \% $ of clusters are estimated to have their cores in this phase
\citep{per98}.
These instabilities can affect several properties, like the $L_X-T_X$ 
relation \citep{fab94,ala98,mar98}.
In addition to cooling flows there is also strong observational 
support for non-gravitational heating of the cluster gas. Simple scaling
arguments predict $L_X \propto T_X^2$ (Kaiser 1986),
while the observed relation satisfies $L_X \propto T_X^3$ \citep{dav93}.
 This discrepancy has been suggested by many
authors \citep{evr91,wu98,pon99} as being evidence for substantial 
heating of the intracluster gas due to energy injection by 
supernova explosions at high redshifts.
Another central question for including star formation in simulations 
of the gaseous component is the growing observational evidence 
for radial gradients in the iron abundance \citep{ez97,fuk98}, 
with possible 
connections to cooling flows \citep{alb98}.

With increasing availability of computational power, numerical hydrodynamical
simulations have attempted to model the effects of radiative 
cooling of the gas in the formation and evolution of cluster galaxies
\citep{kat93,sug98,ann96,yos00,per20,lew20}. 
The numerical problems posed by the inclusion of 
gas cooling are challenging, mainly because the required increase in spatial
resolution also requires that one keeps two-body heating mechanisms
under control.
Moreover, it will be seen that the inclusion of radiative cooling for the gas
cannot be separated from considering also star formation and energy feedback 
from SN explosions, in order to obtain realistic cluster density 
profiles and luminosities.
Previous simulations have produced some conflicting results 
\citep{yos00,per20}, and so the question of the minimum resolution in this
kind of simulation is still to be fully settled.
The purpose of this paper is to test the numerical reliability of SPH
hydrodynamical simulations of cluster formation. These simulations will
include the effects of radiative losses, star formation and energy feedback
from SN. Four different clusters from two cosmological models have been 
studied in several simulations with different numerical inputs and  star
formation modelling. Final profiles are compared in order to assess the 
effects on the integrations
of numerical resolution, and different star formation prescriptions.
The paper is organized as follows. In \S2 I describe the hydrodynamical
simulations with radiative cooling and star formation that have been performed.
The simulation results are then discussed in \S3. In particular, \S3.1 
is dedicated to a comparison between different runs of the final radial density
 and temperature profiles, as well as of the X-ray luminosities.
In \S3.2 the simulation results are compared with previous findings. In \S3.3
simulation runs with different star formation prescriptions are performed
for a chosen test cluster which showed a well defined cooling instability.
In \S3.4 simulation results for runs with different prescriptions for
 star formation are compared against observed data from cooling
flow clusters, in order to assess the consistency with real data 
for the assumed star formation models in the simulations.
Finally the main conclusions are summarized in \S4.

\section{Simulations}
In a previous paper (Valdarnini, Ghizzardi \& Bonometto 1999, hereafter VGB) a 
large set of hydrodynamical simulations was 
used to study global X-ray cluster morphology and its evolution.
The simulations were run using a TREESPH code with no gas cooling or heating.
 In order to assess the numerical reliability of the numerical integrations,
four different clusters were selected as a representative sample of all of the 
simulation clusters.
For this cluster sample, a large set of different integrations was performed 
by varying two numerical input parameters: the number of particles and the 
softening parameter. A comparison between the final 
gas density and temperature profiles, as well as with X-ray luminosities,
then allowed the two-body heating to be kept under control and 
a fairly safe range of allowed values for the numerical parameters to be 
established.
The simulation tests showed the relative importance of different numerical 
effects in these adiabatic simulations. It is therefore natural 
to use the same cluster sample to study the effects of including 
additional physics such as gas cooling and star formation.
A comparison with the previous tests in VGB will show, with 
respect to the adiabatic case, the effects on the final cluster properties 
of considering energy sinks and non-gravitational heating.
Here I give a short description of the cluster simulations performed 
in VGB; the reader is referred to the original paper for more details.
Numerical modelling of gas cooling and gas conversion into collisionless 
stars is described later. 

In VGB three spatially flat cosmological models have been considered.
A standard cold dark matter model (CDM) , a vacuum-energy dominated model
 with $\oml=0.7$ (\La) and a mixed dark matter model (CHDM).
For the Hubble constant $H_0=100 h  Km~ sec^{-1}~ Mpc^{-1}$, 
 $h=0.5$ is used for CDM and CHDM and $h=0.7$ for \La.
For all models, the primeval spectral index $n=1$ and the baryon density
parameter $\Omega_b h^2 = 0.015$. 
For \M ~$\Omega_h = 0.20$ is the HDM density parameter of massive neutrinos;
only one massive species is considered.
All of the models were normalized in order to reproduce the
present cluster abundance \citep{eke96,gir98}. 
 The cosmologies were chosen in order to have simple models
with different properties. 
 For each model an N-body cosmological simulation was first run in a
$L=200h^{-1}Mpc$ comoving box using a $P^3M$ code.
The particle number were $N_p = 10^6$  for the CDM and \M~ models with
$\Omega_m=1$, while $N_p=84^3$ for \La~ with $\Omega_m=0.3$.
The same random numbers were used to set the initial 
conditions for all three cosmological models. 
The simulations started from an initial redshift $z_{in}$.
 At $z=0$ clusters of galaxies were located using a friends--of--friends (FoF) 
algorithm, so as to detect overdensities in excess of $\simeq 200
\Omega_m^{-0.6}$. 
For statistical analysis, VGB selected for each model the 40 most massive
clusters.
For each of these clusters a TREESPH hydrodynamical simulation was 
performed in physical coordinates.
 The integration was accomplished by first locating at $z=0$ the cluster 
center and identifying all of the simulation particles of the cosmological
simulation within $r_{200}$, where the cluster density is 
$\simeq 200\Omega_m^{-0.6}$ times the background density.
These particles are located back at $z_{in}$, in the original simulation box, 
and a cube of size $L_c$ enclosing all of them is then found,  with a size 
$\simeq 15- 25 h^{-1}$ Mpc. 
A lattice of $N_L=22^3$ grid points is set inside this cube;
different lattices were used for each matter component.
At each node position is associated a particle of corresponding mass
 and coordinates.
The particles were then perturbed, using the same random realization 
as for the cosmological simulations.
Additional frequencies are introduced so to sample the higher
Nyquist frequency. The baryon particles are perturbed identically to the CDM 
particles
and their initial temperature is set to $T_i=10^4 \gr$. 
For the TREESPH simulations all the particles which lie inside 
a sphere of radius $L_c/2$ are kept.
External gravitational fields are modelled by considering a larger
cube of side $2L_c$, inside the cube particle
positions are set as for the smaller cube, but with no gas.
The number of grid points is the same as for the inner cube, so that
masses are 8 times larger than those of the inner cube; 
particles of the larger cube are considered only outside the smaller one.
After the particle positions are perturbed, only those within a sphere of 
radius 
$L_c$ from the cluster center are kept for the TREEPSH simulations.
 This multimass grid technique  has already been used in cluster simulations
 by Katz \& White (1993) and Navarro, Frenk and White (1995).
 
For each particle, the gravitational softening parameters 
 are set according to the scaling $\varepsilon_i \propto m_i^{1/3}$.
The numerical integrations were performed with a tolerance parameter 
$\theta=0.7$, without quadrupole corrections.
The reliability of the numerical resolution was tested in VGB by taking the most
massive and least massive clusters (labels $00$ and $39$, respectively) for
the two models \M ~and \La. For each of them, different numerical 
tests were carried out. Accordingly to VGB, the cluster simulations can be 
performed with an adequate resolution using a number of gas particles 
$N_g \simgt 5,000$ and a gas softening parameter $\varepsilon_g \simlt 50
h^{-1} Kpc$. For the collisionless component the corresponding values 
are scaled accordingly.
In Table~\ref{tab:ta1} the reference values for the four clusters are given.
If one introduces radiative cooling then the final values for the core 
radius of the gas density are expected to be smaller than in the no-cooling
case.
This implies that the numerical simulations must have smaller values for the gas
softening $\varepsilon_g$, which in SPH sets a minimum resolvable scale,  
since SPH smoothing lengths are constrained to be 
$ \simgt \varepsilon_g/4$. 
The shape of the density profile implies, with respect to the adiabatic case,
higher values for the gas and dark matter densities at the core radius.
This in turn implies that larger values for the number of particles
are required in simulations of cooling clusters.
This is essential in order to reduce the values of particle masses and 
hence the two-body heating time $\tau_r$, which 
approximately scales as $\tau_r \propto 1/(\rho_d m_d)$, where ~$\rho_d$ and $m_d$ 
are the dark matter density and particle mass.

In order to check for these numerical effects, for each of the four test 
clusters 
a set of five TREESPH simulations was performed,
 using the same initial  conditions and with the inclusion of radiative
cooling, but for different values of $\varepsilon_g$ and $N_g$. 
With respect to the reference case, the other matter components have their
particle numbers changed  in proportion to the change in $N_g$.
For the cluster \La$00$, Table ~\ref{tab:ntesta} reports the values of $\varepsilon_g$ and $N_g$ 
for the five simulation runs. Table~\ref{tab:ntestc} is for \M$00$. For these 
two clusters the generic simulation
has cluster index cl$00-j$, with $j=00,01,...,05$. The cluster cl$00-00$ is
the reference case without cooling. The simulations have been carried out with 
the same values for the other numerical parameters that were used in VGB, with 
the difference that here the minimum allowed time step for gas particles is 
$\Delta t_m=6.9~10^5 yr$.
For the clusters \La$39$ and \M$39$, the parameters of the numerical tests 
cl$39-j$ are 
given in Table ~\ref{tab:ntestd}. For these simulation runs, the number of particles
and the initial redshifts are the same as for the $00$ clusters. Therefore 
Table ~\ref{tab:ntestd} reports only the particle masses and softenings.
The effects of radiative cooling are modelled in these simulations by 
adding to the SPH thermal energy equation an energy-sink term  
(Hernquist \& Katz 1989, eq. [2.29]). The total cooling function includes 
contributions from
 recombination and collisional excitation, bremsstrahlung and inverse
Compton cooling. For cluster temperatures which satisfy 
$ T \simgt 2 KeV$, 
 a condition which is always satisfied for the cluster sample studied here, 
the dominant cooling mechanisms are free-free 
transitions, but line cooling becomes important for small clusters and groups.
For the same reason, heating from an ionizing UV background is not included 
in the thermal energy equation.
The radiative cooling is computed for a gas having primordial abundances
$X=0.75$, $Y=0.25$ with zero metallicity.
In the simulation runs where the gas particles are allowed to convert part
of their mass into stars (see below), the back effects of metallicities 
on the cooling function are neglected. 
This is a valid approximation as long as $ T \simgt 2 KeV$ and stellar 
metallicities are below $ Z \simlt Z_{\odot}~( [Fe/H] \simlt 0) $ 
\citep{bry98,car98}.

 In addition to these simulations, which include the effects of radiative
cooling, a mirror simulation was performed for each of them.
The mirror runs had an additional prescription which allowed eligible gas to be
 turned into stars.
These simulations are indexed as cl$00-k$ and cl$39-k$, where 
$k=j+5$ and $j$ is the index of the pure cooling runs in 
Tables~\ref{tab:ntesta},
\ref{tab:ntestc} \& ~\ref{tab:ntestd}.
For the cluster \La$00$ the simulation parameters for the cooling runs with
star formation, cl$00-k$, are given in Table~\ref{tab:ntestb}.
The numerical parameters of the simulations with radiative cooling and star
formation are the same as the corresponding cooling simulations.
For this reason Table~\ref{tab:ntestb} reports parameters only for the 
cluster \La$00$.
The additional simulation with $k=11$ in Table ~\ref{tab:ntestb} has the same 
parameters as the $k=10$ run, but with $\theta=1$ and
quadrupole corrections, instead of $\theta=0.7$.
This is done in order to check
the accuracy of the gravitational integration when a collisionless 
population, with a different distribution from that of dark matter, is 
added to the simulation.
Allowing the gas to cool radiatively will produce dense clumps of gas
at low temperatures ($\simeq 10^4 \gr $). Thus cooling times will 
become shorter and even denser regions will develop. This is known as 
the overcooling instability  \citep{sug98}. In these regions the gas
will be thermally unstable and will meet the physical 
conditions to form stars.
In TREESPH simulations, star formation (SF) processes have been implemented 
using two different algorithms \citep{kat92,na93}. According 
to Katz (1992), a gas particle is in a star forming region if the flow
is convergent and the local sound crossing time is larger 
than the dynamical time (i.e. the fluid is Jeans unstable). 
These two conditions read 

\be
\left\{ \brr {ll}
\nabla \cdot \vec v_i <&0  \\ 
h_i/c_i >& \sqrt{3 \pi/16G\rho_i}\equiv \tau_d ,
\err
\right .
\label{eq:sfc}
\ee
where $\vec v_i$ is the particle velocity, $h_i$ the SPH 
smoothing length and $c_i$ is the local sound velocity.
In a more refined version, Katz, Weinberg \& Hernquist (1996, hereafter KWH) 
introduced two additional requirements: a star forming region must have a 
minimum physical hydrogen number density $n_H=0.1 cm^{-3}$ and the local 
gas density must satisfy $ \rho_g / \bar \rho_g > 55.7 $ ( this follows 
from an isothermal profile giving a mean virialized overdensity of 
$\simeq 169 $). 
If a gas particle meets these criteria then it is selected as an eligible 
particle to form stars.
In regions where the gas density is depressed because of gravitational 
softening, the Jeans criterion is not applied, in order to avoid an 
underestimate of
the local star formation rate (SFR) (Katz 1992, eq.[2]).
The local SFR obeys the equation 
 \be 
 d\rho_{g}/dt=-c_{\star} \rho_g /\tau_g= -d \rho_{\star} /dt~,
\label{eq:rg}
\ee
where $\rho_g$ is the gas density, $\rho_{\star}$ is the  
star density , $c_{\star}$ is a characteristic dimensionless efficiency 
parameter, $\tau_g$ is the local collapse time (the maximum 
of the local cooling time $\tau_c$ and the dynamical time $\tau_d$). 
Gas particles 
with $ T \simlt 10^4 \gr $ have long cooling times and $\tau_g=\tau_d$.
The probability that a gas particle will form stars in a time step $\Delta t$ 
is given by 
 \be 
 p=1-\exp {(-c_{\star} \Delta t /\tau_g)}.
\label{eq:ps}
\ee
A uniform random number $\xi_r$ is generated at every time step for each
of the gas particles satisfying the star formation criterion, and 
 equation (\ref{eq:ps}) is used to compute  the formation probability $p$. 
If $\xi_r < p$ then a mass fraction
$\varepsilon_{\star}$ of the mass of the gas is converted into a new 
collisionless particle. This star particle has the position, velocity and
gravitational softening of the original gas particle. 
Typical assumed values are  $\eps=1/3 $ and $c_{\star}=0.1$ (KWH).

The second algorithm for implementing SF in TREESPH simulations 
has been introduced by Navarro \& White (1993 , hereafter NW). 
According to NW,
 any gas particle which is in a convergent flow and for which the density 
exceeds a threshold , i.e. 

\be
\left\{ \brr {ll}
\nabla \cdot \vec v_i <&0  \\ 
\rho_g ~~>&\rho_{g,c}=7 \cdot 10^{-26} gr cm ^{-3},
\err
\right .
\label{eq:sfd}
\ee

will have cooling time shorter than the dynamical time
 and will soon cool to $ T \simlt 10^4 \gr $ , thus satisfying 
the Jeans instability criterion. The two conditions (\ref{eq:sfd}) are  
necessary and sufficient conditions for selecting gas particles as prone to SF.
For the local SFR, NW adopted equation ({\ref{eq:rg}) with $c_{\star}=1$, 
$\tau_g= \tau_d$ and $\eps=1/2$ as the condition for which a gas particle can 
convert part of its mass into 
a star particle. These two algorithms will be referred to hereafter as
KWH and NW, respectively.

The numerical tests cl$00-k$ and cl$39-k$ have been performed 
following the  NW prescription for selecting gas particles 
which can form stars.
The NW method has been preferred over KWH because of its simpler assumptions
about the physical conditions of the gas in star forming regions.
Because of the many physical processes involved in SF, having a minimal number
 of
assumptions can reduce possible biases in hydrodynamical simulations
when modelling the local SFR.
For a single representative cluster, a detailed comparison has been made
between the cluster properties obtained using the two methods and different
input values for the SF parameters.
These simulation runs with numerical modelling of SF also include 
energy feedback to cluster gas from supernova (SN) explosions.
Once a star particle is created it can release energy into the
 surrounding gas through SN explosions. This energy is converted
 into heat of the neighboring gas  at each time step, according to the stellar
 lifetime and initial mass function. A standard Miller-Scalo (1979) mass 
function has been adopted in the mass range from $0.1$ to $100 M_{\odot}$.
 All of the stars with masses above $ 8 \msu$ end as  SNe, leaving a $1.4 \msu$
remnant.
Each SN explosion produces $\varepsilon_{SN} \simeq 10^{51} $ erg , or 
$\simeq 7.5\cdot 10^{48}$ erg $/\msu $, which is added to the thermal energy of the gas.
Current time steps are much smaller than stellar lifetimes, and so the SN energy
is released gradually into the gas according to the lifetime of stars 
of different masses. 
At each time step, the fraction of stars 
releasing their energy into the medium 
is calculated for any star particle  
and the corresponding SN energy 
is spread over neighboring gas particles according to the SPH 
smoothing prescription.
SN explosions also inject enriched material into the intracluster medium,
thus increasing its metallicity with time.
According to Steinmetz \& Muller (1994) $ p_Z=0.357m -2.2$ solar masses 
of heavy elements are synthesized by a SN progenitor of mass $m$. 
The enrichment in metals of the intracluster medium is modelled 
as follows \citep{sm94,car98}. 
Each SPH gas particle initially has zero metallicity; star 
particles are produced with the metallicity of the parent gas particle at 
the epoch of their creation.
The metallicity of gas particles is successively enriched at each time step 
according to the fraction of exploding SNe associated with each star particle.
The mass in metals produced by these explosions is calculated in 
accordance with the specified function $p_Z$, and is 
added to the metallicities of the gas neighbors of the star particle. 
This mass is distributed over the neighbors using a smoothing procedure 
identical to that implemented for spreading the SN feedback energy  
of star particles among internal energies of the  gas neighbors. 
According to the same procedure, the current mass fraction of exploding 
SNe with $M \geq 8 \msu $ is also added to the mass of the 
 gas particles, with the exception of the $1.4 \msu$ remnant.

\section{Results}
\subsection{Cluster simulations with radiative cooling and cooling plus star
formation }
The radial density and temperature profiles for the pure cooling runs are 
shown in Figures~\ref{fig:l1a} \& \ref{fig:m1a}. The cluster center has been identified as the maximum of
the gas density. For each radial bin spherical averaged quantities have been 
obtained by estimating hydrodynamical variables at $100$ grid points uniformly spaced in angular coordinates.
Densities and temperatures at the grid points were computed from SPH variables 
according to the SPH smoothing procedure.
The cluster \La$00$ is a particularly neat example of the effects at work in 
the simulations. As can be inferred from the softening values reported 
in Table~\ref{tab:ntesta}, the numerical strategy has been first to run cl$00-00$ with 
the additional cooling prescriptions (cl$00-01$) and in subsequent 
runs the value of $\epg$ has been reduced in order to resolve the core
radius of the gas density profile. Figure 1a shows that this is 
not achieved: whatever is the value of $\epg$, there is no evidence of 
a gas core radius, the gas density continues to rise steeply at the cluster 
center without any indication of converging to a constant value.
This result is in strict agreement with those of others 
 \citep{sug98,ann96,per20,lew20} and it is known as the overcooling instability;
at the cluster center cooling times are very short because the gas density is 
high, thus drawing in more more material.
The fact that the gas central density continues to rise as the spatial 
resolution is increased suggests that the extent of the (physical) effect 
is limited by the numerical resolution of the simulation.
Of the four test clusters, \M$00$ is the only one which does not 
show the cooling instability with the exception of cl$00-05$ (Figure 2a).
The reason for this behavior may be dynamical. 
Buote \& Tsai (1996) measured  a negative correlation of the cooling
rate with the cluster X-ray substructure.
The cooling instability can then be 
strongly suppressed when the cluster is still in a young dynamical state,
with a large fraction of substructure.
For the cosmology considered, VGB found that the clusters studied in the \M~ 
model with $\omm=1$ 
had many more substructures than those in the low-density \La~model. 
In this case \M$00$ could be marginally stable, with instability being triggered
by numerical effects when the central value of the gas density increases 
because the simulation resolution is increased. 

In the simulation runs the number of particles was increased 
as $\epg$ was reduced so as to keep 2-body heating under control. 
The relaxation time $\tau_r$ , due to 2-body effects, is defined as
\bea 
\tau_r = & 0.34 \frac {\sigma_1^3}{G^2 m_d \rho_d \ln \Lambda} \simeq 
6.7 \cdot 10^5 Gyr  \left ( \frac {\sigma_1 }{10^3 Km sec^{-1}} \right )^3 
\nonumber \\
 & \times \frac { h^{-2} } { (m_d /10^{11} M_{\odot})} \frac {1} {(\rho_d / \rho_c) 
\ln \Lambda }~,
\label{eq:tr}
\eea
where $\sigma_1$ is the 1-D dark matter velocity dispersion, $G$ is the 
gravitational constant, $\rho_d$ is the dark matter density; $\ln \Lambda$ 
is the Coulomb logarithm, with $\Lambda \simeq R_h/ 4 \epg$, and $R_h$ 
is the half-mass radius. Typical values are $ \ln \Lambda \simeq 3$;
standard theory gives $\Lambda \simeq R_h/ \varepsilon$,   
the factor $4$ above accounts for the softening bimodal distribution 
(Farouki \& Salpeter 1982).
For the simulations cl$00-00$ and cl$39-00$ 
the relaxation time $\tau_r$ has been estimated at radius  
$\simeq 0.05 r_{200} \simeq 100 Kpc$,  
approximately the resolution limit of these simulations.
 At this length scale 
$\rho_d / \rho_c \simeq 2 \cdot 10^4 $. The values of $\tau_r$ range
from $ \simeq 13 Gyr $ (cl$00-00$) to $ \simeq 17 Gyr$ (cl$39-00$).
For the simulation runs with increased resolution the minimum resolvable
scale is set by $\epg$, and $\rho /\rho_c \simeq 5 \cdot 10^4$ at a 
fiducial scale 
$ \simeq 50 Kpc$. These rather high values of $\rho_d$  
are a consequence of the slightly steeper profiles for dark matter in 
the inner regions, with respect to the case with no cooling.
For these runs, the increase in $\rho_d$ is compensated by a corresponding 
increase in the particle number and hence a smaller value for $m_d$, so that 
$\tau_r$ is approximately constant at the scale considered and is 
close to the Hubble time. 
Another timescale which is relevant  
for these simulations is the cooling time $\tau_c$, defined as

\be 
\tau_c =  \frac {3}{2}  \frac {n k_B T }{\Lambda_c} ~ ,
\label{eq:tc}
\ee
where  $k_B$ is the Boltzmann constant, $n$ is the gas number density,
$T$ is the gas temperature and $\Lambda_c\simeq 5.2 \cdot 10^{-28} T^{1/2} n^2 
erg sec^{-1} cm^{-3}$ is the cooling function.
In the central gas regions $\tau_c << H_0^{-1}$ and a cooling instability will 
develop. In Figure \ref{fig:tc} $\tau_c$ is shown as a function of radius for the four test 
clusters. For the cooling runs cl$00-05$, cl$39-05$ $\tau_c$ is always below 
$\simeq 20 Gyr$ for $ r \simlt 100 Kpc$, with the exception of \La$39$ ,
which has a bump but then a strong fall in $\tau_c$ proceeding inwards.
For the sake of reference, $\tau_c$ for the no-cooling case ( $-00$) has also
 been plotted. 
Because of the presence of a gas core radius in this case 
$\tau_c$ approaches a constant value towards the center.
According to Steinmetz \& White (1997) gas cooling will be 
affected by artificial 2-body heating  unless $\tau_c(r) < \tau_r(r)$. 
This condition is satisfied if the dark particle mass is smaller 
than the critical value 
\be 
M_c =  2 \cdot 10^9   T_6  f_{0.05} M_{\odot},
\label{eq:mc}
\ee
where  $T_6$ is the the gas temperature in units of $10^6 \gr$, $f_{0.05}$ is 
the ratio $f=\rho_g / \rho_d$ in units of $0.05$. 
For the simulated clusters studied here $ T_6 \simeq 50-100$,
$f_{0.05}\simgt 0.5$   for $ r \simlt 100Kpc$ and $M_c$ is 
always above $m_d$. The simulations can then be considered 
free from numerical effects which can dominate the gas behavior.
In the cooling simulations  this condition might be violated near the cluster 
center, where $T_6 \simlt 10$   for $ r \simlt 50 Kpc$.   
However, in these regions $ \tau_c << \tau_d \simeq 27 h ^{-1} Gyr / 
\sqrt {\rho_d /\rho_c}$ and the cooling is effective in removing the gas 
energy at a faster rate than the one set by dynamical effects.

The temperature profiles show a decrease for  $ r \simlt 100 Kpc$ and a 
drastic drop in the central values, where cooling is most effective.
This inversion in temperature takes place in all of the tests considered.
 Peak values for the gas temperature are located at $ \simeq 100 Kpc$
 ($ 0.05-0.1$ of $r_{200}$). Between this distance from the cluster center and 
 $r_{200}$ the gas temperature decreases with radius and the clusters clearly 
cannot be considered isothermal.
These results are in agreement with those of  Pearce et al. (2000),
 and suggest that the global cluster properties are affected by cooling 
processes
active on inner scales , where the cooling time is short 
(see also Lewis et al. 2000).
The most important cluster variable which is affected by these results 
is the cluster X-ray luminosity. For evaluating $L_X$ the standard SPH 
estimator gives   
\be 
L_X = 5.2 \cdot 10^{-28} \frac {1} { (\mu m_p)^2 } \sum_i^{N_g} m_i \rho_i T_i^{1/2} erg sec^{-1},
\label{eq:lx}
\ee
where, for the cooling function, the bremsstrahlung emissivity has been 
approximated with a gaunt factor of $1.2$ (eq.[2] of Suginohara \& Ostriker 
1998), 
$\mu=0.6$ is the mean molecular weight, and $m_p$ is the proton mass. 
The summation is over all of the gas particles within $r_{200}$. 
Figure \ref{fig:lx}
shows the behavior of $L_X$  at $z=0$ as a function of $\epg$  
for the pure cooling runs (open symbols).
There is not a clear convergence of $L_X$ as the resolution is increased. 
In fact $L_X$ obeys the approximate scaling  $ L_X \propto 1/\epg^{\sim 5}$.
Similar results for  $L_X$ have been obtained by Anninos \& Norman (1996)
in their convergence study of simulations of X-ray clusters.

The unphysically high values found for $L_X$ in the pure cooling runs arise
because the gas density continues to increase steadily  at the cluster center,
while the conditions of high gas density and low temperature cause the 
gas to become Jeans unstable.
Thus the treatment of gas cooling in cluster simulations cannot be 
decoupled  from a modelling of the physical processes turning the cold, 
dense, gas into stars.
The cooling simulations have therefore been rerun with the inclusion
in the integrations of an algorithm for converting gas into stars.  
This used the NW method, with parameters $c_{\star}=1$ and
$\eps=1/2$. In these simulations gas particles can produce star particles
without any limit on the number of star-forming events. 
Furthermore, the SN explosion
energies and metallicities that star particles can produce are smoothed 
over 
$32$ gas neighbors but with an upper limit of $h_M=15Kpc$ for the SPH smoothing
length. This is in order to avoid unphysical heating of the gas over length 
scales
 much larger than those involved in the SF activities.
This upper limit is also justified by the lack of diffusion in the ICM
of the metals injected from galaxies \citep{ez97}.
The results obtained are shown in Figures~\ref{fig:l2a} and \ref{fig:m2a}; the index of the simulations
is $k=j+5$, where $j$ is the index of the cooling runs.
The most important result is that the inclusion of an SF model has been 
effective in removing the unphysical gas behavior, which now shows a 
well defined core radius in the radial density plots.
This is valid in all of the cases considered, with the expection of \M$00$.
The simulation runs cl$00-k$ do not show, for this cluster, evidence of an SF 
activity, a result which is in agreement with the lack of a cooling instability 
in the cooling simulations cl$00-j$. The shapes of the temperature profiles
show that convergence is achieved for $N_g \simgt 20,000$.
For the simulation runs with the highest resolution, 
all of the central values for the gas temperatures at $r=10Kpc$ are  
within a factor $\simeq 1.5$. 
\M$39$ is an exception to this rule, with cl$39-10$ still not showing 
a flat temperature profile in the inner regions and resembling that of 
cl$39-05$.
It is unlikely that the source of the discrepancy is due to a convergence 
problem: cl$39-11$ has a slightly larger accuracy  
in the computation of the gravitational forces (Hernquist 1987), 
nevertheless its temperature
profile has these discrepancies largely removed.
For \M$39$ the peculiarity of cl$39-10$ in the final temperature
profile, with respect to the other numerical tests shown in Figure 4b,
could be of a numerical nature: for a certain accuracy in the tree evaluation of
the gravitational forces, matter subclumps might form during the integrations
which can then modify the gas dynamics.
The formation of these sub-clumps is triggered by the approximations 
involved in the truncation of the multipole expansion of the cluster 
gravitational potential.
The statistical occurrence of this effect should be small, because it is not 
observed in the other three test clusters.
For a tree method the errors involved in the multipole expansion of
the gravitational potential have been estimated by Hernquist (1987) assuming
a spherical, isotropic Plummer model for the mass distribution of
$N$ test particles. A comparison against the accelerations obtained by a
direct sum shows that in the large $N$ limit ($N \simgt 30,000$) the errors
in the tree evaluation of the accelerations are negligible for a monopole
expansion if $\theta \simlt 1$. The inclusion of the quadrupole terms 
improves the accuracy of the force computation, for $\theta\simeq1$ the errors
in the forces are those of a monopole expansion with $\theta\simeq0.8$.
The convergence in $T(r)$ obtained for cl$39-11$ suggests that for
a given accuracy, quadrupole corrections should be preferred when 
evaluating tree forces.

The radial density profiles of the star component
are also shown  
for the various runs  in the density plots.  
The slope of these profiles is approximately $\simeq -3$, a value 
close to  the one observed for galaxy populations in galaxy 
clusters. 
In all of the simulations, the gas density profiles have a well-defined 
core radius, with size $r_c \simeq 50-100 Kpc$, approximately $0.05$ 
of the virial radii. From the density profiles note also that the
gas core radii are smaller than in the no-cooling runs, outwards of $r_c$ the 
density profiles are very similar to the no-cooling cases. 
The temperature profiles 
increase inwards from the virial radius up to $ \simeq 100-200 Kpc$. 
Thereafter the profiles stay almost flat, or with a modest decrease in
 $T(r)$ towards $r=0$. The strong drop of the temperatures in the very
central regions for the cooling runs is no longer seen, the inclusion of
a star formation prescription having been effective in removing the 
cold gas particles ($\simlt 10^4\gr$) from the cluster centers.

Cooling timescales $\tau_c(r)$ are plotted in Figure \ref{fig:tc}; 
the dashed line in the four panels is for the simulation runs including SF.
As a general rule, for each test cluster, the $\tau_c$ are almost 
indistinguishable
in all of the simulation tests for $r\simgt 100Kpc$. 
In the simulations including SF, $\tau_c$ moves 
toward the no-cooling case in the cluster inner regions because of the 
reduction in the gas central density.
The central values of $\tau_c$ are well below the present age of the 
universe for all of the models,
with the exception of \M$00$. This cluster does not show a cooling instability
and has $\tau_c \simeq 20 Gyr$ in the cluster core.
Accretion rates $\dot{M}(r) =4 \pi \rho_g r^2 v_r$ are plotted in 
Figure~\ref{fig:md}
for the same test clusters as in Figure~\ref{fig:tc}. For each radial bin, 
spherical averages of $\dot M(r) $ are shown only for negative 
values of $v_r$. For the test clusters in the $\Lambda$-dominated 
cosmology there is a well defined radial infall of matter within
$ r \simlt 100Kpc$, compared to the adiabatic run. 
These inflows of matter can be compared with those estimated from
X-ray data for cooling flow clusters. For example Thomas, Fabian 
\& Nulsen (1987) present mass deposition profiles 
 $\dot M(<r) $  for a sample of 11 cooling flow clusters.
They obtain for Abell 478 $\dot M \simeq 10^3 \msu yr^{-1}$ at 
$r \simeq 300 Kpc$ (Fig. 7 of their paper). For this cluster they quote a 
measured line-of-sight velocity
dispersion of $ \simeq 750 Km sec^{-1}$ and the estimated cluster virial mass
 ( $\simeq 4\cdot 10^{14} h^{-1} \msu$)  is close to that of $\Lambda$CDM39.    
This accretion rate is in good agreement with the values shown in 
Figure~\ref{fig:md} at $r \simeq 300 Kpc$  for $\Lambda$CDM39 in the 
simulation including SF.
Note that in order to compare $\dot M(<r)$  with  $\dot M(r) $  one 
is implicitly assuming a steady-state.
 The clusters in the cosmology with $\Omega_m=1$ show values of $\dot M$ 
which are higher
in the adiabatic case, in comparison with the simulation runs including 
cooling and star formation. The reason for this discrepancy 
is of a dynamical nature: \M$00$, for example, has a large radial infall 
velocity because it is still out of equilibrium, with material collapsing
onto the center.
These results seem to support the hypothesis of an anti-correlation
between the strength of the cooling flow and the dynamical state of the
cluster, as measured by the power ratios, for example.
Buote \& Tsai (1996) demonstrated the existence of 
such an anti-correlation, as expected from a dependence of the 
cooling flow rate on the cluster dynamical state. A statistical 
analysis of this correlation is beyond the scope of this paper and
is left to future work, where the analysis will be performed 
for the whole sample of $40$ clusters for each of the three cosmological
models (VGB).
 A striking result is the convergence of the final X-ray luminosities
for the simulation runs including SF for the four test clusters.
In Figure~\ref{fig:lx} the filled symbols refer to these simulations.
The divergence for $\epg \rightarrow 0 $ of the cooling runs is completely
removed and the plotted values are quite stable. An exception is 
cl$39-10$ of \M$39$ (the black square in the bottom right panel) which has 
$L_X \simeq 5 \cdot 10^{44} erg sec^{-1}$. As previously discussed, the 
peculiarity of this cluster is of a dynamical nature and there is not a 
question of
convergence of hydrodynamical variables. In fact cl$39-11$ has $L_X \simeq 2 
\cdot 10^{44} erg sec^{-1}$, a value in full agreement with the values of the 
other runs. Compared with the non-radiative case, the luminosities 
are stable (\M) or increase by a factor $\simeq 2$  (\La).

\subsection{Comparison with previous simulations}

These results can be compared with previous findings from hydrodynamical 
cluster simulations, in which the gas is allowed to undergo cooling and star 
formation or is subject to a prescription for the treatment of cold 
particles.  
The density plots can be compared to analogous plots shown in Lewis et al. 
(2000).  In their paper Lewis et al. analyzed five simulations of a 
cluster with $M_{200} \simeq 4 \cdot 10^{14} M_{\odot}$ in a CDM universe 
with $\Omega_m=1$ and $h=0.5$.
Of these simulations, the $Cool+SF$ allows gas to undergo cooling 
and star formation, using the KWH method with $\cs=0.1$ and 
 $\varepsilon_{SN}=1$. 
Although the cosmologies are different, a rough comparison can be made 
for \La$39$, which has the closest virial mass to their test cluster. 
For \La$39$ the simulations cl$39-10$ or cl$39-11$ have a numerical
resolution comparable to the Lewis et al. (2000) simulations 
(compare Tables ~\ref{tab:ntestb} \& ~\ref{tab:ntestc} with Table 2 of Lewis et al.).
Thus the density plots in Figure 3b can be compared with Figure 7 of Lewis 
et al. . The results are encouraging, there is a rough agreement 
for the various baryonic components, although in Lewis et al. 
there is a central spike in the gas density which is not observed
in the present simulations.
 A substantial difference is instead found for the temperature profiles:
all of the simulation runs show a tendency for $T(r)$ to recover
an almost flat behavior in the inner regions (with the exception of
cl$39-10$ for \M$39$, which is peculiar for the reasons 
previously outlined). Lewis et al. found instead that $T(r)$ in their 
$Cool+SF$ simulation reaches a peak value of $ \simeq 8\cdot 10^7 \gr$ within
$ \simeq 40 Kpc$ (Figure 9 of Lewis et al.).
The radial behaviors of Figure~\ref{fig:tc} ($\tau_c(r)$) compare well 
with Figure 8 of Lewis et al. (2000).

Hydrodynamical simulations of cooling clusters have also been analyzed by
Pearce et al. (2000) and Yoshikawa, Jing \& Suto (2000).
I discuss here in detail how the results of \S3.1 compare with the
cluster properties of Yoshikawa, Jing \& Suto (2000, hereafter YJS).
The simulated clusters of the Pearce et al. (2000) runs have  
 properties similar to the YJS clusters, the only substantial differences 
being found for the X-ray luminosities.

In their paper YJS analyze results 
from a set of cosmological SPH simulations and concluded that estimates of 
the X-ray luminosities
are biased by the numerical resolution of the simulations and are not 
reliable. 
The cosmological simulations of  YJS 
were performed in a flat, $\Lambda$-dominated CDM cosmology. The values of
the background cosmological parameter are identical to the ones 
chosen here for the cluster simulations in the \La~model.
They performed simulations with two different box sizes : $L=75 h^{-1}Mpc$ and
$L=150 h^{-1}Mpc$. They took $128^3$  gas  particles, an equal number of 
dark particles, and a comoving softening parameter 
$\varepsilon=L/1280$.
In the simulations including radiative cooling, cold gas particles were removed 
from the summations defining local gas variables if the Jeans condition 
in equation (\ref{eq:sfc}) is satisfied 
(simulation runs with label UJ in their notation). 
This constraint is slightly different from the one of Pearce et al. (2000), 
but is almost identical for $T_i \simlt 10 ^4 \gr$.
Their prescription is phenomenological and is intended to take into account 
the process of galaxy formation.
For the cosmological simulation with box size $L=150 h^{-1}Mpc$, the 
numerical resolution for the most luminous clusters is comparable to that 
of the cl$00-11$ run.
The most massive cluster in the simulation with label 150UJ has 
$\simeq 1.56 \cdot 10^{15} M_{\odot}$, a value close to that of
$M_{200}$ in Table~\ref{tab:ta1} for \La$00$.
For this cluster YJS found at $z=0$ a
bolometric X-ray luminosity $\simeq 2 \cdot 10^{45} erg sec^{-1}$,
 about twice the corresponding  value of $L_X$ for cl$00-11$.
A comparison between the cluster density and temperature profiles
 (Figure 1 of their paper, central panel) and the 
analogous plots of the present paper (cl$00-11$ in Figure 3) shows that there 
are 
substantial differences at $r\simlt 50Kpc$.
In YJS the cluster gas density shows no evidence
for a core radius; the mass-weighted temperature profile has an inversion at 
the cluster
center, as also found by  Pearce et al. (2000), while cl$00-11$ shows 
a nearly flat profile for $T(r)$. These differences in the gas profiles
at the cluster centers are probably due to the different methods employed
for the treatment of the cooled gas, rather than to the numerical
resolution of the simulations. 

YJS adopted the phenomenological prescription of treating separately those gas
particles which satisfy the Jeans criterion.
Therefore the gas distribution consists of gas particles in a `hot' X-ray
emitting phase together with a population of cold particles at temperatures
around $10^4 \gr$, the latter being localized at the cluster center.
The temperature profiles in YJS include the cold gas population  
\footnote{ K. Yoshikawa: private communication} and show
a steep decrease at the cluster center. In the simulation run cl$00-11$ 
this is not observed because the star formation prescription has been effective 
in converting the cooled gas at the cluster center into the form of stars.
For the simulation runs with cooling and star formation, the 
mass-weighted
temperatures $T_m(sim)$ of the clusters
\La$00$ and \La$39$ can be compared with those expected from the isothermal
mass-temperature relation. At the present epoch, this relation reads:

\be 
k_B T_X \simeq 5.2 \gamma \left( \frac{\Omega_m \Delta_c}{178 }\right)^{1/3}
\left( \frac {M}{10^{15}h^{-1} \msu}\right )^{2/3} KeV,
\label{eq:tx}
\ee
where $M$ is the cluster virial mass, $\Delta_c\simeq 178\Omega_m^{-0.55}$ is 
the virialized cluster
overdensity and $\gamma$ is a fudge factor.
YJS found that the mass-weighted temperatures of their simulated clusters 
are well fitted by equation \ref{eq:tx} with $\gamma=1.2$.
For the cluster \La$00$ (\La$39$) I find from the 
simulations cl$00-11$(cl$39-11$) that $T_m(sim)=5.6(3.1) KeV$, while the theoretical relation 
(\ref{eq:tx}) gives
$T_m(th)=5.9(2.8) KeV$. Thus the 
 mass-weighted  temperatures of the simulated clusters are in close
agreement with the theoretical predictions and also with those found
by YJS in their simulations.
 
As previously shown in the simulation runs including star formation, 
the X-ray luminosities are found to be numerically stable and converging
to reliable values. With respect to the adiabatic runs, the $L_X$ are found 
to be constant or with a modest increase (Figure~\ref{fig:lx}).
These results are at variance with those of Pearce et al. (2000), who found
 that cooling clusters are less luminous than those in the no-cooling runs. 
Similar results  have been obtained by YJS and Lewis et al. (2000), who measured
 an increase in the final X-ray luminosity when cooling was included.
A comparison with the Cool+SF simulation of Lewis et al. (2000) is 
difficult because of the different cosmologies and cluster virial masses. 
YJS analyzed the $L_X-T_X$ relation 
 obtained from the simulated clusters in the UJ test runs at $z=0$,
 from the simulations with two different box sizes. 
The cluster luminosities in the simulation with $L=150 h^{-1}Mpc$ 
are found to be underestimated with respect to the ones of the 
$L=75 h^{-1}Mpc$ simulation box. 
YJS draw the conclusion that, in order to
have reliable cluster luminosities, simulations of cooling clusters 
require a much higher numerical resolution than the one employed in
their simulations.
In the simulations with $L=150 h^{-1}Mpc$, the mass of the gas particles 
is $\simeq 2\cdot 10^{10}M_{\odot}$, a factor $\simeq 2 $ larger than that 
of cl$00-10$ for \La$00$. Therefore their conclusion seems to be in conflict 
with what is claimed here, that substantial convergence  
 in X-ray luminosities is achieved for the highest resolution 
simulation runs.
The source of this discrepancy lies in the different numerical
approaches. 
YJS simulated cluster evolution in a
cosmological box with a constant gas particle mass.
As outlined by the authors, the resolution problem is severe for
the less luminous clusters in the simulation box.
The multimass technique described in \S2 is instead used here
to simulate a single cluster. The results of the numerical simulations
show that convergence in the gas variables is obtained for 
each {\it single } cluster whenever $N_g \simgt 20,000$.
A comparison with the numerical resolution adopted by YJS is useful. 
In their simulations with $L=150 h^{-1}Mpc$  the gas particle mass is 
comparable to 
the value found here for which \La$00$ can be safely analyzed.
However they have a constant mass resolution 
and their cluster sample has a lower limit of $M> 10^{14} M_{\odot}$.
An application of the numerical parameters required here to a cluster 
with a virial mass  of $\simeq 10^{14} M_{\odot}$ would give a gas particle 
 mass of $\simeq \Omega_b 10^{14} /( \Omega_m 2.2  \cdot 10^4) 
\simeq  5 \cdot 10^8 M_{\odot}$, a factor $\simeq 30$ smaller than 
that being used in the $L=150 h^{-1}Mpc$ simulations.
These values also show that cosmological simulations require a 
number of gas particles $ \simeq (400 L /150 h^{-1} Mpc)^3 $ in order to 
give realistic estimates for the statistical properties of X-ray clusters.

\subsection{ Simulations with different star formation prescriptions}
The numerical tests studied here give the range of numerical parameters 
for which the results of cooling cluster simulations, including
SF according to the NW prescription, reach numerical convergence and can be 
considered stable.
A different question concerning the reliability of the numerical results
is the sensitivity of the estimated cluster properties to the 
numerical method used to describe star formation and energy feedback from 
stars in the hydrodynamical simulations.
To investigate this, final results for different simulations have been compared 
for a single test cluster (\La$00$). The simulations were performed 
with the same numerical parameters as for the cl$00-11$ run, but using
different SF methods or parameters. This was done in order to  demonstrate 
the effects of the SF modelling on the gas variables.
A summary of these simulations with different SF prescriptions 
is reported in Table ~\ref{tab:ts1}.
The simulation with index I is the standard case with which 
previous cooling+SF runs have been performed.
Thus this simulation just corresponds to cl$00-11$.
Two simulation runs correspond to the NW method but with a 
different feedback energy for SN explosions (I and V).
In the other three runs the KWH prescription was adopted for
converting gas particles into stars. Two of them (II and III) compare the 
results obtained for a different star formation efficiency parameter
$c_{\star}$, with the other parameters being held fixed;
in a third run (IV) the fraction $\eps$ of mass of the gas 
converted to stars is varied.
The results of the different approaches are shown in Figures~\ref{fig:sf}
\& \ref{fig:mg}. For the simulation V, only the star formation rate (SFR)
and the X-ray luminosity versus time have been plotted in the two left panels
 of Figure~\ref{fig:sf}.
This is because the simulation V produced final results almost identical
to the reference case I.
In simulation V the feedback SN energy was set to
 $\varepsilon_{SN} = 10^{50} erg $, a value ten times smaller than
that of the simulation run I.
A comparison between the results plotted in Figure~\ref{fig:sf}
shows that the two simulations have an identical evolution for the X-ray 
luminosity, but a different SFR. This is because of the smaller 
amount of SN feedback energy which is added to the thermal energy
of the gas in the run V. This in turn implies higher gas densities 
and SF rates for simulation V. The differences become negligible 
for $t \simgt 8 Gyr$. The final gas profiles, as well as the other 
variables, are identical.

The simulations with the KWH method and different $\eps$ ( II and IV),
give similar results and show that the choice of the mass fraction 
$\eps$ is not important in modelling the star formation processes.
The most important differences are found between the KWH  simulations
with different $c_{\star}$ (II and IV). The differences are
dramatic in the final X-ray luminosities, which differ by a factor 
$ \simeq 40$.
The source of this discrepancy lies in the different gas density profiles,
which have substantial differences in the cluster core regions for
$ r \simlt 100 Kpc$. These differences are localized at the cluster center;
beyond $ r \simeq 100 Kpc$ all of the profiles converge, as  
shown in the plots of Figure~\ref{fig:sf}.
The temperature profiles have a peak value of $ \simeq 10^8 \gr$ at 
$ r \simeq 100Kpc$ and thereafter decline outwards by a factor $\sim 2$ 
out to $r_{200}$. Below $\sim 100Kpc $ the profiles instead show  
large differences. Compared to run I (NW) the two simulations 
with $\cs=0.1$ have gas temperatures which decrease inwards by a 
factor $\sim 10$ from $\sim 100Kpc$ down to $r\sim 10Kpc$.
These radial decays follow because of the less efficient conversion
of the cooled gas into stars compared to the NW run.
These results can be compared with those of Lewis et al. (2000),
who used the same star formation method and parameters, although in 
a different cosmology. The temperature profile of their Cool+SF 
simulation shows a similar decrease outside of the cluster core
(Fig.9 of their paper). Is is difficult to compare the radial runs at $ r < 
100Kpc$ because the authors adopt a linear scale for the plots.
The gas density profile of run II rises steeply towards the cluster center
from $ r \simeq 100Kpc$; a similar behavior is shown by Lewis et al.
(2000) for their hot gas population ( see their Fig.7 ). 
The simulation run with $\cs=1$ has  radial profiles
much closer (but not identical) to the NW ones. These differences are also
reflected in the estimated emission-weighted temperatures (see 
Table~\ref{tab:ts1}). For
simulations I and III $T_{em} \simeq 8-9KeV$, while $T_{em} \simeq 2 KeV$ 
for run II. Note that for this simulation most of the contribution to 
$T_{em}$ comes from within the $50Kpc$ cluster core.

There is a remarkable agreement for the ratio of the cluster mass locked into
 stars to the gas mass, which is $ \simeq 10 \%$  at $r_{200}$ for all of the 
runs considered  (Figure~\ref{fig:mg}, bottom left panel), 
with respect to the observational values ( Evrard 1997).
Also the density profiles of the stars produced are very similar.
As a general rule, one can say that the global cluster properties 
are not affected by the SF prescription adopted.
The source of the higher gas densities in the cluster core for 
the simulations with $c_{\star}=0.1$, compared to what is 
found for $c_{\star}=1$ , is in the different SFR that the
two simulations use during the integration.
In the top left panel of Figure~\ref{fig:sf} the SF rates are 
plotted for the different runs.
The rates are plotted versus time instead of redshift because 
otherwise final differences would have been compressed.
It can be seen that the simulations II and IV have a  different SF 
histories, with respect to runs III and I.
At early times runs II and IV have a lower SFR than the 
simulation with $c_{\star}=1$. This behavior is reversed after 
$t \simeq 10 Gyr~(z \simeq  0.3)$,
with simulations II and IV showing a substantial SF activity, while
runs III  and I have a sharp decline in their SF rates.
As final results, the two simulations II and IV have less gas 
converted to stars in the cluster core.

According to KWH, in a simulation with $\cs=0.1$, the collapsing gas will reach
an equilibrium density higher than in the one with $\cs=1$.
This is because of the reduced efficiency in the conversion of the gas into
 stars for the simulation with $\cs=0.1$, which implies that 
gas collapse will proceed until pressure from the thermal
energy of the gas is able to prevent further gas collapse, reaching an 
equilibrium at higher gas densities.
This in turn implies higher SF rates so that, after an initial transient,
simulations with $\cs=0.1$ will have higher gas densities and SF rates 
compared to the $\cs=1$ runs.
This expected behavior is what is found in the plots of Figure~\ref{fig:sf}, 
the NW simulation giving identical results to the KWH run with $\cs=1$.
The bottom right plot of Figure~\ref{fig:mg} shows $M_s(z)$ the cluster mass 
in stars at the redshift z and is particularly useful 
in analyzing differences in the SF histories between different
runs.
The mass in stars for the NW run stops growing at $z=0.5$, and stays
flat until $z=0$. The KWH simulation with $\cs=1$ has a similar  trend
but with a somewhat steeper slope in the growth of $M_s(z)$.
An important result is that the final values of the cluster mass 
in stars vary by only $\simeq 20 \%$ from $M_s\simeq 1.5 \cdot 10^{13}
M_{\odot}$. As previously outlined, this shows the reliability of different
SF prescriptions in reproducing global cluster properties.
In the simulation results, differences between the methods  
 arise in the SF rates, which are determined  by different choices
of the star formation efficiency parameter $\cs$.
In Figure~\ref{fig:mg} for simulations II and IV, the values of $M_s(z)$ at high
redshifts are below the corresponding values of the $\cs=1$ runs.
For these simulations with $\cs=0.1$, $M_s(z)$ has a continuous growth 
and is smaller than $M_s(z)$ of the NW run until $z\simeq 0.2$.
Note that the final masses in stars for the different runs do not correspond
to what would naively be expected from the different shapes of the
 gas density profiles in the cluster core,
 $M_s(z=0)$ being determined by the overall SF history.
The SF rates of the simulations with $\cs=0.1$ have the 
expected behavior, with respect to those of the $\cs=1$ runs, but  
the high resolution runs of the next section will show that 
these SF rates are depressed by the 
numerical resolution of the simulation.

The emission weighted metallicity profiles at $z=0$ are shown in 
Figure~\ref{fig:met}. 
Conversion from the mass fraction 
in metals associated with each particle to the iron mass was
accomplished according to the relationship between the 
total metallicity and the iron abundance $[Fe/H]$ of
Tantalo, Chiosi \& Bressan (1998).
The relative iron abundances $Z=Fe/H$ in the radial bins were estimated 
from the individual iron mass of each gas particle using the 
SPH smoothing procedure. 
These profiles were still rather noisy, and to obtain the final  
radial profiles of iron abundance, a further smoothing was performed by considering 
only five distinct radial bins and averaging over neighboring bins.
An important result for the metallicity profiles is that  
the simulation runs show the existence of radial gradients,
with decreasing metallicities as the radius increases.
This is in broad agreement with observational data 
for cooling flow clusters 
(Ezawa et al. 1997; Kikuci et al. 1999; Buote 2000; White 2000; 
De Grandi \& Molendi 2001).
For the simulation runs I and III  
$Z\simgt 0.5 Z_{\odot}$ at the cluster center
and ~ $Z \simlt 0.1 Z_{\odot}$
at $r\simgt 100Kpc$~, 
with a value for the solar abundance 
$Z_{\odot}=(Fe/H)_{\odot}\equiv4.68\cdot 10^{-5}$ \citep{an89}.
At $r\simgt 100Kpc$ these values are $\sim 1/3$ of the measured abundances  
obtained for a sample of 9 cooling flow clusters \citep{de01}.
This deficit of iron abundance is probably due to the lack of metal enrichment
of the intracluster medium from SN of type Ia and will be analyzed
in a future paper, where the metallicity
dependence of the cooling function will be implemented in the simulations.
The KWH runs with $\cs=0.1$ have metallicity
profiles with abundances well below the NW run, this is because of 
the different SF histories,  
 with simulations II and IV having a lower SFR  
 at early times ($\simlt 10 Gyr)$.
The KWH run with $\cs=1$ has, on average, a steeper metallicity gradient
than the NW run; here again this is because of the 
 different SF histories, as shown by the different growth of
 $M_s(z)$.  

\subsubsection{ High resolution simulations with different star formation prescriptions}
An important point about the simulation results related to the SF parameters, 
such as for example the SFR, is their dependence 
on the numerical resolution adopted.
The numerical tests of Table~\ref{tab:ts1} have been performed with the same number 
of particles and
softening parameters (cl$00-11$ in Table ~\ref{tab:ntestb}), and for the 
particle gas mass $m_g \simeq 1.2 \cdot 10^{10} M_{\odot}$.
The resolution of the mass of gas is clearly inadequate for modelling a single 
galaxy formation process. 
SPH simulations of galaxy formation processes have been debated by many 
authors (Thacker \& Couchman 2000, and references cited therein).
In order to resolve the internal dynamics and to follow shock 
evolution of a forming galaxy a minimum of 
 $\simeq 10^4 $ particles is required (Thacker et al. 2000).
Lia, Carraro \& Salucci (2000) discussed the dependence of the 
mass resolution on the SFR in gas-dynamical simulations of a collapsing 
spheroid.
According to their results, the total mass of gas converted into stars
diminishes from $90\%$ to $\simeq 80\%$  of the initial mass of gas, passing 
from the low resolution run with $ 2,000$ particles to the high 
resolution run with more than $ \simeq 10^4$ particles.
However, early simulations (Evrard 1988, NW) showed that SPH simulations with 
even a small number of particles  ($\simeq 100$) can converge to stable results 
as far as global properties are concerned.
For example Figure 20 of NW indicates that the final mass of stars in their SPH 
simulations of a rotating cloud is already close to the convergence value
for a number of gas particles $N_{g} \simgt 100$.

It is therefore important to assess the effects of numerical resolution on 
final results  in the simulations with different SF prescriptions.
This has been done by running again simulations I, II and III of Table~
\ref{tab:ts1}  
but 
with a number of particles increased by a factor $ \simeq 3$.
For the gas particles $N_g =69,599$,$~m_g=2.4 \cdot 10^9 h^{-1} M_{\odot}$ and 
$\epg=10.5 h^{-1} Kpc$. The cold dark matter particles have these values 
scaled in proportion. These simulations will be referred as IH, IIH and IIIH,
respectively. The numerical parameters are those of cl$00-11$H in Table 
~\ref{tab:ntestb}.
Different SF parameters correspond to the ones of Table ~\ref{tab:ts1}.  
The simulation results are shown in Figure~\ref{fig:sft}. For simulations
IH there are not appreciable differences in the radial profiles.
This confirms the results of \S3.1, that the NW runs have reached 
numerical convergence in the physical variables for the numerical
parameters of the cl$00-10$ simulation of Table ~\ref{tab:ntestb}. The 
profiles of 
simulation IIH are instead different from those of run II at 
$r \simlt 50 Kpc$. The strong drop in $T(r)$ has been removed and 
the gas density profile is much closer to the NW one. 
Simulation IIIH gives final profiles very similar to the ones of 
the parent simulation.
The bottom left panel of Figure~\ref{fig:sft} shows that high resolution
runs have final X-ray luminosities which can differ by a factor $\sim 2$
from the parent simulations.

SF rates are shown in the the top left panel 
 of Figure~\ref{fig:sft} and at $ z>0$ there are
 large differences between the high resolution run IIH and the 
parent simulation.
For $t \simgt 5 Gyr$, simulations IH and IIIH give the best performances, with
the SF rates closely following the ones of the lower resolution runs.  
Note the strong decline in the SFR after $z=0.3$~.
At early times ($t\simlt 5Gyr$) the SF rates of simulations I and III are
much lower than those of the corresponding high resolution runs. For these
runs, the peak of the SF activity is shifted from $z \simeq 0.7$ up to 
$z\simeq 2$ ($t \simeq 3Gyr$). This shows that in order to correctly sample
the SF rate over the whole cluster evolution the numerical resolution must be 
at least that of the high resolution runs.
For $t \simgt 5Gyr$, simulations I and III 
have cluster SF rates in good agreement with those of the 
corresponding high resolution runs,   
while simulation II does not show that convergence is achieved when 
the resolution is increased.
For simulations I and III, the X-ray variables are not affected by the 
undersampling of the SF rates at early times; for example, the X-ray
luminosities (Figure~\ref{fig:sft},~bottom left panel)
are fairly stable under an increase in the numerical resolution.

The results of these high resolution simulations show that in the 
large $N_g$ regime, different star formation
methods approach similar gas profiles. For the simulations of 
Table~\ref{tab:ts1}, the 
differences found between the profiles of the runs with $\cs=0.1$ and the 
others arise because in the former case, with respect to the KWH run with 
$\cs=1$, the increase in the SF rate 
corresponding to a decrease in the value of $\cs$ 
is limited by the numerical resolution.
The upper left panel of Figure ~\ref{fig:sft} shows that for the high
resolution run IIH the SF rate is much higher than in simulation
II.

From the simulation results it follows that the NW 
 method is the most efficient in the removal of cold gas from the cluster 
center. As already stressed, in principle all the methods 
converge to the same profiles  when $N_{gas}$ gets very large. 
The runs with $c_{\star}=0.1$ are the ones with the most important
differences in the temperature profiles, compared to the other runs, and the 
reasons of these
differences have been previously discussed.
 An explanation of why the NW method is so efficient relies on the 
 chosen  criterion for selecting cold gas subject to star formation, 
 the method being based on a density threshold. If the local gas 
density $\rho_i$ exceeds this threshold ($7. 10^{-26} gr cm ^{-3}$) then a mass
fraction $\varepsilon_{\star}=1/2$ of the gas
particle is converted into a star particle. In very high density regions, the 
timescales of star formation are very short and the SF
process removes the gas very quickly.
Because the criterion is based on a
  density threshold this means that \emph {all} the gas particles above this
threshold are selected for a star forming event. 

Differences in the temperature profiles between the NW and KWH runs with
 $c_{\star}=1$ are minimal and are localized at the cluster center, 
these differences being due to the different dependence of the 
two methods on the resolution limit of the simulations.
The two criteria of SF depend on the  numerical resolution of the simulations
through the
SPH smoothing lengths $h_i$, which determine $\rho_i$ and are 
constrained by the lower limit $h_i \simgt \varepsilon_{g}/4 \equiv h_{min}$, 
where $\varepsilon_{g}$ is the gravitational softening length.  
As the gas density increases toward the cluster center, the 
 smoothing lengths $h_i$ get smaller, until they reach the limit
$h_{min}$.
The results of the simulations show that at the cluster center, 
in the small value regime $h_i \rightarrow h_{min}$, 
the NW criterion for identifying cold gas particles is less sensitive to
the resolution limit $h_{min}$ than the KWH  method based on the local
Jeans instability.

 To summarize, the above results demonstrate that simulations I and III 
of Table~\ref{tab:ts1} have an adequate numerical resolution to reliably 
predict 
X-ray cluster properties, such as the X-ray luminosity.
For simulation II ( KWH  with $\cs=0.1$ ) there are still differences 
at the cluster core between the final profiles when the numerical 
resolution is increased. Run IIH has a final $L_X$ which is very large 
compared to the expected range of values from the luminosity-temperature
relation (see below).
For the simulation runs I and III $L_X \simeq 10^{45} erg sec^{-1}$,
while $L_X \simeq 4 \cdot 10^{46} erg sec^{-1}$ for the runs with $\cs=0.1$.
These large discrepancies can be reduced by adopting the phenomenological
prescription of removing from the summation (\ref{eq:lx}), cold gas particles 
with temperatures below a cut-off value $T_c$.
For the runs II and IV with $\cs=0.1$, $L_X \simeq 10^{45} erg sec^{-1}$
if the above prescription is adopted with $T_c \simeq 3\cdot 10^7 \gr$.
It is important to note that the NW run has an $L_X$ that shows no sign of 
evolution for $ t \simgt 5 Gyr$ ( $ z \simlt 1.2$ ).

\subsection{ Comparison with observational parameters}
These differences between different methods suggest that a reliable SF 
algorithm should be chosen by requiring that simulation results should 
consistently
satisfy a wide set of different observational constraints on cluster properties.
To this end, the results of the simulation runs reported in Table~\ref{tab:ts1}
 have been 
compared with several cluster observations.
The data investigated are : the $L_X-T_X$ relation for cooling flow
clusters and the estimated SFR in rich clusters at the present epoch.  
These observations have been chosen because the plots of Figure~\ref{fig:sf}
showed large differences between simulated clusters for the variables connected with these data.

For cooling flow clusters, Allen \& Fabian (1998a) have studied the relation 
between the bolometric X-ray luminosity $L_{bX}$ and the cluster temperature.
They use $ASCA$ spectra and $ROSAT$ images to construct a sample of 30 
luminous clusters ($ L_{bX} > 10 ^{45} erg sec^{-1}$), 
21 of which have central cooling
times $ < 10 Gyr$ and are identified as cooling flow (CF) clusters.
Their best-fit relationships are derived for a cosmology with 
$\Omega_m=1$ and $h=0.5$, thus the values reported in Table~\ref{tab:ta1} of 
their
paper must  be rescaled to those for a flat cosmology 
with $\Omega_m=0.3$ and $h=0.7$~.
Allen \& Fabian (1998)  fit the $ L_{bX}-T_X$ relation with a power-law 
of the form $kT_X=PL_{bX}^Q$, where $kT_X$ is the spectral fit temperature 
of a single isothermal model in $KeV$ and $L_{bX}$ is in units
of $10^{44} erg sec^{-1}$.
The data values used are those of model C in Table~\ref{tab:ta1}, which 
according to the
authors gives the best $\chi^2$ values.
Thus the coefficients $P$ and $Q$ obtained here for the $\Lambda$-cosmology
are those corresponding in Allen \& Fabian (1998) to the $\chi^2$  
only for the entry CFs in their Table 2.
The least-square estimator gives $P=1.97 \pm 0.33$ and $Q=0.42 \pm 0.05$,
error bars are the one-parameter $68\% $ confidence limits, with $\chi^2/d.o.f=1.03$ and a 
mean gas fraction 
at $360 Kpc$ of $ f_g =0.11 \pm 0.03$.
For the simulation runs I and III 
 $ L_{bX} \simeq 1.3 \cdot 10^{45} erg sec^{-1} $  
and the predicted cluster temperature is $T^{pred}_X=5.8 \pm 1.2 eV$.
The error bar represents the  $1\sigma$ confidence  interval.
The simulated cluster in the NW run has an emission-weighted temperature 
$T_X^{em}=8.3 KeV$ and a mass-weighted temperature $T_X^m =5.6 KeV$.
These values are consistent with those extracted from Figure 5 of
 YJS for the most massive clusters in their UJ simulations. 
They are also in good agreement with the the values 
predicted from the mass-temperature relation inferred 
from cluster $ASCA$ and $ROSAT$ data \citep{ho99,ne20}.
For the simulation run III the values of $ L_{bX}$  and the weighted 
temperatures
 are similar to the ones of the NW simulation. 
Mathiesen \& Evrard (2001) have discussed how measurements of the intracluster
medium spectral fit temperature $T_s$ are related to $T_m$ and showed
that $T_s$ can be considered as a nearly unbiased estimator of $T_m$.
Thus the values of the mass-weighted temperatures reported in 
Table~\ref{tab:ts1}  can be 
compared with the corresponding $T^{pred}_X$. 
The simulations I and III are consistent at the 
$1\sigma$  
level with the $ L_{bX}-T_X$ relation derived from a sample of cooling flow 
clusters. Simulations II and IV have 
$ L_{bX} \simeq 4. \cdot 10^{46} erg sec^{-1} $  and the predicted cluster temperature is outside the $2\sigma$ limits.

Another observational test is to compare for the different simulation runs 
the rates of SF with the observed mean SFR 
 of cluster galaxies.
There is now a strong observational evidence that star formation 
 in a cluster environment is strongly suppressed 
with respect that of field galaxies \citep{bal98,po99}.
Kodama \& Bower (2000) have estimated the SF histories 
in rich cluster cores for four clusters, using photometric models
constructed from 7 CNOC clusters at $0.23 < z<0.43$, and found
a strong decline in the SF rate relative to the field for $z <1$.
The SF histories of the four clusters are shown in Figure 9 
of Kodama \& Bower (2000). One of these clusters is Coma,
for which the measured values for the mass in galaxies, hot gas 
and cluster total mass \citep{wha93} are in the same range 
as those of the simulated cluster \La$00$. Thus a comparison between the SFR 
estimated from cluster simulations and the observed data can be 
made consistently.
Although the numerical resolution of the simulations I-III is inadequate to
correctly sample the SFR of the galaxies in the cluster, 
the previous discussion about simulations IH, IIH and IIIH has shown
that even for the low-resolution runs of Table~\ref{tab:ts1}, the global cluster
SFR is roughly reproduced.
Therefore the SF rates computed in simulations I-III can be used
for making a crude comparison with the values derived from  
Kodama \& Bower (2000), in order to estimate the 
consistency of the simulation results for different SF algorithms
with observations.

The SF rates plotted in Figure~\ref{fig:sf} have been estimated by summing 
the mass of gas converted into stars 
in time bins of size $\Delta t_s=3 \cdot 10^7 yr$. 
For simulations I and III, at $z=0$ $\Delta t_s << \tau_g$ 
and the random process used to sample the distribution (\ref{eq:ps}) can be
approximated as Poissonian, thus the
SFR $ \simeq 70 \pm 70 h^{-2} M_{\odot} yr ^{-1}$; in fact there is only
one single event in the last four time bins. For simulation II,  
the condition  $\Delta t_s >> \tau_g$  applies and the 
SFR $\simeq 330 \pm 110 h^{-2} M_{\odot} yr ^{-1}$; the dispersion
has been computed over the last four bins.
These values must be rescaled to the normalization adopted by 
Kodama \& Bower (2000), who reported the integrated rates for galaxies 
in the cluster cores  in units of $10^{12} M_{\odot}$ within a 
radius enclosing $1/3$ of the galaxy population.
In the simulations of Table ~\ref{tab:ts1}
 $M_s\simeq 1.5 \cdot 10^{13}$ is the total mass in stars , with a small 
scatter between different runs. Thus 
SFR$_{(KB)}$ $ \simeq 15 \pm 15 h^{-2} M_{\odot} yr ^{-1} /10^{12} M_{\odot} $ 
for simulations I and III, while 
SFR$_{(KB)}$ $ \simeq 66 \pm 22 h^{-2} M_{\odot} yr ^{-1} /10^{12} M_{\odot} $ 
for simulation run II.
These values can be compared with SFR$_{(Coma)}$ 
$ \simeq 0.25  h^{-2} M_{\odot} yr ^{-1} /10^{12} M_{\odot} $ 
found by Kodama \& Bower (2000) for the Coma cluster at $z\simeq0.1$.
For the cosmology chosen by Kodama \& Bower ($\Omega_m=0.2$ and $h=0.5$ 
without a cosmological constant), at this redshift the  Coma cluster
has an age comparable to that of the simulated clusters at $z=0$.
Therefore  simulations I and III have a cluster SFR which is 
consistent with the observed values. The simulation II, with the 
KWH method and $c_{\star}=0.1$, has instead an SFR  clearly above the 
observed limits.
For the high resolution runs, the cluster SF rates are above the observed limits
for Coma, with simulation IH being marginally consistent at a 
$2\sigma$ level.
The Coma cluster has been chosen for comparison because it has a measured 
cluster SFR and the estimated values for the mass components match those 
of \La$00$. 
However the Coma cluster does not show a cooling flow activity and the 
cluster SF rates of the simulations would have been much smaller if 
\La$00$ did not have a cooling instability.

\section{Conclusions}
 In this paper I have analyzed how the gas and X-ray properties of clusters 
of galaxies estimated from hydrodynamical SPH simulations are affected when
radiative cooling is included. 
It has been found that in order to get reasonable results the inclusion of
cooling cannot be decoupled from a prescription  for  converting cold
gas particles into stars. 
The final results depend on the star formation prescription adopted and
the numerical resolution.
When the cooling gas particles are converted into stars according to the NW
prescription,
 the final cluster profiles are found to be remarkably stable under changes in
the numerical resolution.

This is achieved by using for the individual SPH cluster 
simulations a number of gas particles 
$N_g \simgt 20,000$ and a gas softening parameter $\varepsilon_g \simlt 20
h^{-1} Kpc$. 
Above this numerical threshold, estimates of the final cluster properties 
can differ among different runs by a factor $\simlt 2$.
It must be stressed that these conclusions are valid for the cluster 
sample studied here. As discussed in \S2, numerical simulations of clusters 
less massive than those of Table~\ref{tab:ta1} must take into 
account the metallicity dependence of the cooling function.
In this case, the convergence of X-ray variables may require a numerical
resolution higher than those considered here.

If the KWH 
star formation prescription  is adopted, with a star formation efficiency 
parameter $\cs=1$, the stability of the final results with respect to the 
numerical resolution of the simulation is satisfied for the same range of numerical parameters
given above for the runs with the NW prescription.
The KWH simulations with $\cs=0.1$ have been found to give final results 
which are much more dependent on the simulation numerical resolution.
For the KWH runs, the final differences in the gas density profiles at 
$r \simlt 100Kpc$ are 
a consequence of the different SF histories, which depend on $\cs$.
These differences strongly affect the X-ray luminosity $L_X$, which is 
then the simulation variable most sensitive to the value of $\cs$.
The results demonstrate that, for the same simulated cluster,
 different SF algorithms yield final gas distributions with
differences localized at the cluster center.
Global cluster properties, such as the total mass in stars, are robust to 
different SF prescriptions, while X-ray luminosities can differ by large
factors.

A relevant difference for the simulation runs with the NW prescriptions, 
with respect to previous simulations,  is 
the flatness of the profiles. For the \La$00$ cluster the gas profile shows 
a core radius of
$r_c\simeq 20kpc$ and the temperature profile is almost flat for
$r \simlt 100 Kpc$. 
This is at variance with what expected for a cluster with a cooling flow,
 but the \La$00$ cluster has at its center a cooling time $\tau_c(0)\simeq 1/3$
of the universe age $t_U \simeq 13.5 Gyr$. The other two clusters which
experienced a cooling flow (\La$39$ and \M$39$), have $\tau_c(0) \simeq
1 Gyr $ and temperature profiles which decline by a factor $ \sim 2 $ within the
$\sim 200Kpc$ cluster central regions.

A comparison of the temperature profiles with the results of Pearce et al. 
(2000)
and YJS is problematic because these authors adopt a phenomenological
prescription for removing cold gas particles from SPH estimates.
In the case of YJS the temperature profiles include the cold gas population 
and show a steep decline at the cluster centers.  
The profiles of the Cool+SF simulation of Lewis et al. (2000) can be compared
with those of the corresponding KWH run with $\cs=0.1$ and do not show 
inconsistencies. Therefore the profiles obtained for the NW run are not
inconsistent with previous simulations.
The temperature profiles show a decline between the cluster central regions
and the virial radii, but  the observational
evidence for temperature gradients is 
controversial \citep{marb98,ir99,whi20}.

To summarize, SPH hydrodynamical simulations of clusters of galaxies with 
radiative cooling and suitable star formation algorithms have been proved
to be numerically stable, giving cluster X-ray properties which satisfy a
set of observational constraints. 
 An application of the numerical schemes adopted to a simulated cluster sample
can be used  to reliably predict the evolution of the cluster X-ray 
luminosity and temperature function in different cosmological models. 
X-ray cluster surveys from the $XMM$ mission can then provide strong 
constraints on the allowed cosmological background parameters, by comparing
cluster data with simulation results.

\clearpage

\begin{figure}[tbph]
 \begin{center}
  \includegraphics[width=15cm,keepaspectratio]{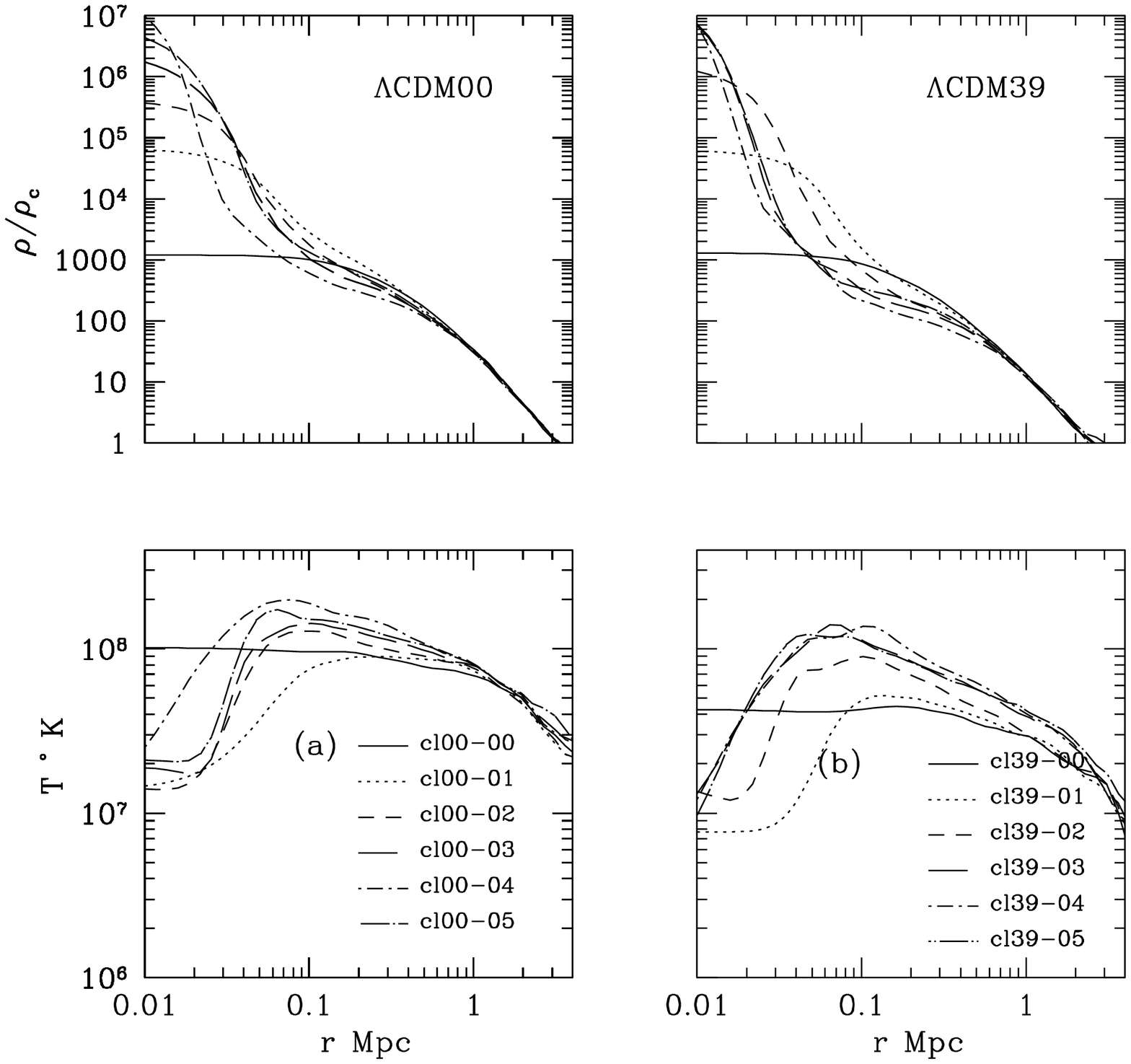} 
 \end{center}
\figcaption[Fig1.eps]{
Radial dependence of the gas density and temperature profile at $z=0$ in 
simulations with radiative cooling.
The left panel is for the test cluster \La$00$, the right panel for \La$39$.
 In each panel, the upper plot is for densities
 and the lower plot is for gas temperatures. The density is in units 
of the critical density.
 The different curves are for integrations with different numbers of  
 particles  and different softening parameters (see Tables 2 \& 3).
The simulation run with index $-00$ is the integration without cooling.
 \label{fig:l1a}}
\end{figure}

\clearpage
\begin{figure}[tbph]
 \begin{center}
  \includegraphics[width=15cm,keepaspectratio]{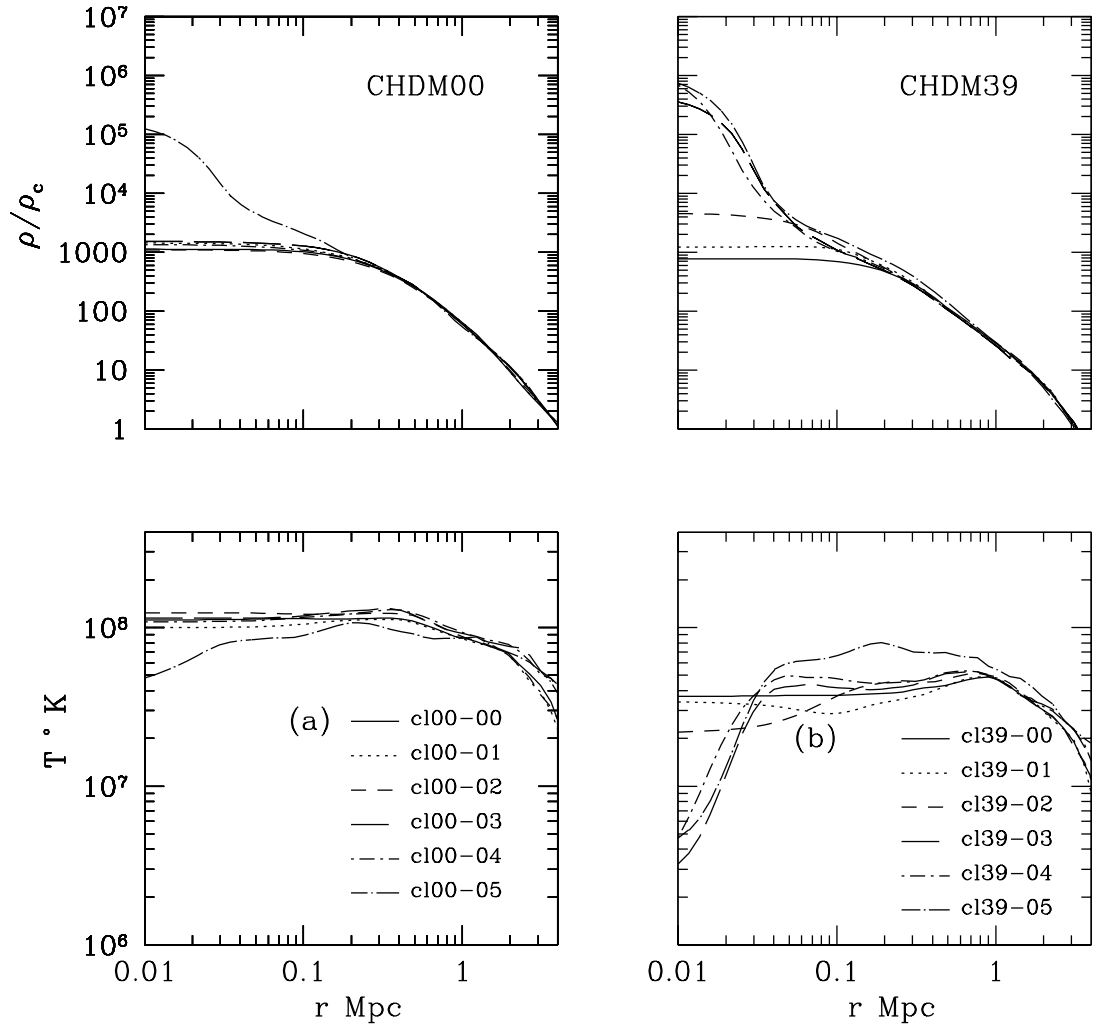} 
 \end{center}
\figcaption[Fig2.eps]{
As in Fig. \ref{fig:l1a}, but for the test clusters \M$00$ and \M$39$.
The numerical parameters of the simulations are 
given in Tables ~\ref{tab:ntestc} and \ref{tab:ntestd}.
 \label{fig:m1a}}
\end{figure}

\clearpage
\begin{figure}[tbph]
 \begin{center}
  \includegraphics[width=15cm,keepaspectratio]{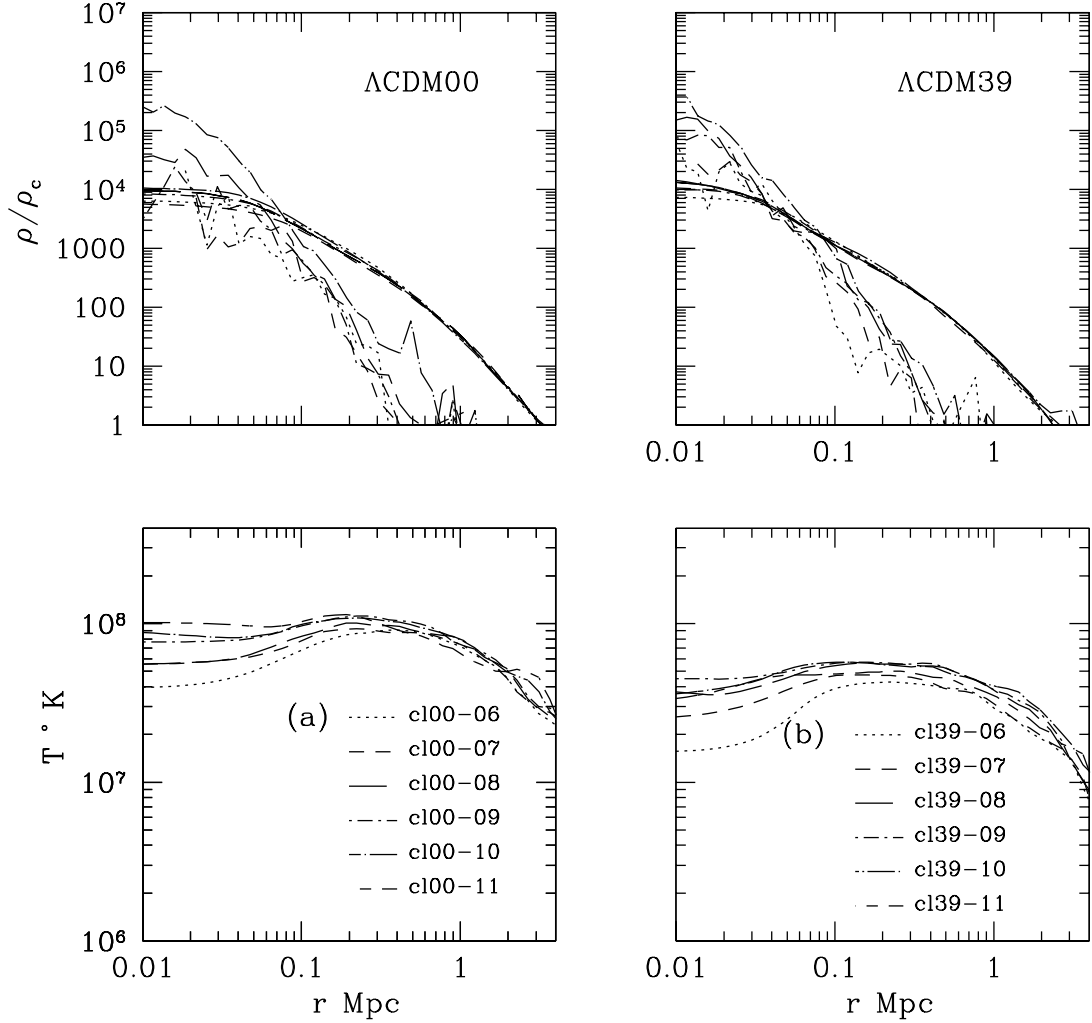} 
 \end{center}
\figcaption[Fig3.eps]{
 Final density and temperature profiles for the simulation 
runs including radiative cooling and star formation. The gas is allowed 
to convert into stars according to the NW star formation prescription 
(see text).
The simulations are the same shown in Fig.~\ref{fig:l1a}, with the numerical 
parameters reported in Tables 3 \& 5. The index of the simulations is 
cl$00-k$ or cl$39-k$, with $k=j+5$ and $j=1,5$ is the index of the cooling runs 
in Tables 2 \& 5. The simulations with index $k=11$ have the same 
parameters as the $k=10$ runs, but with the gravitational tolerance parameter 
$\theta =1$ and quadrupole corrections enabled. The other simulations 
were performed with $\theta=0.7$ without quadrupole corrections.
In the density plots, the lines with a steeper slope ($\simeq -3$) show the 
density behavior of the star component. \label{fig:l2a}}
\end{figure}

\clearpage
\begin{figure}[tbph]
 \begin{center}
  \includegraphics[width=15cm,keepaspectratio]{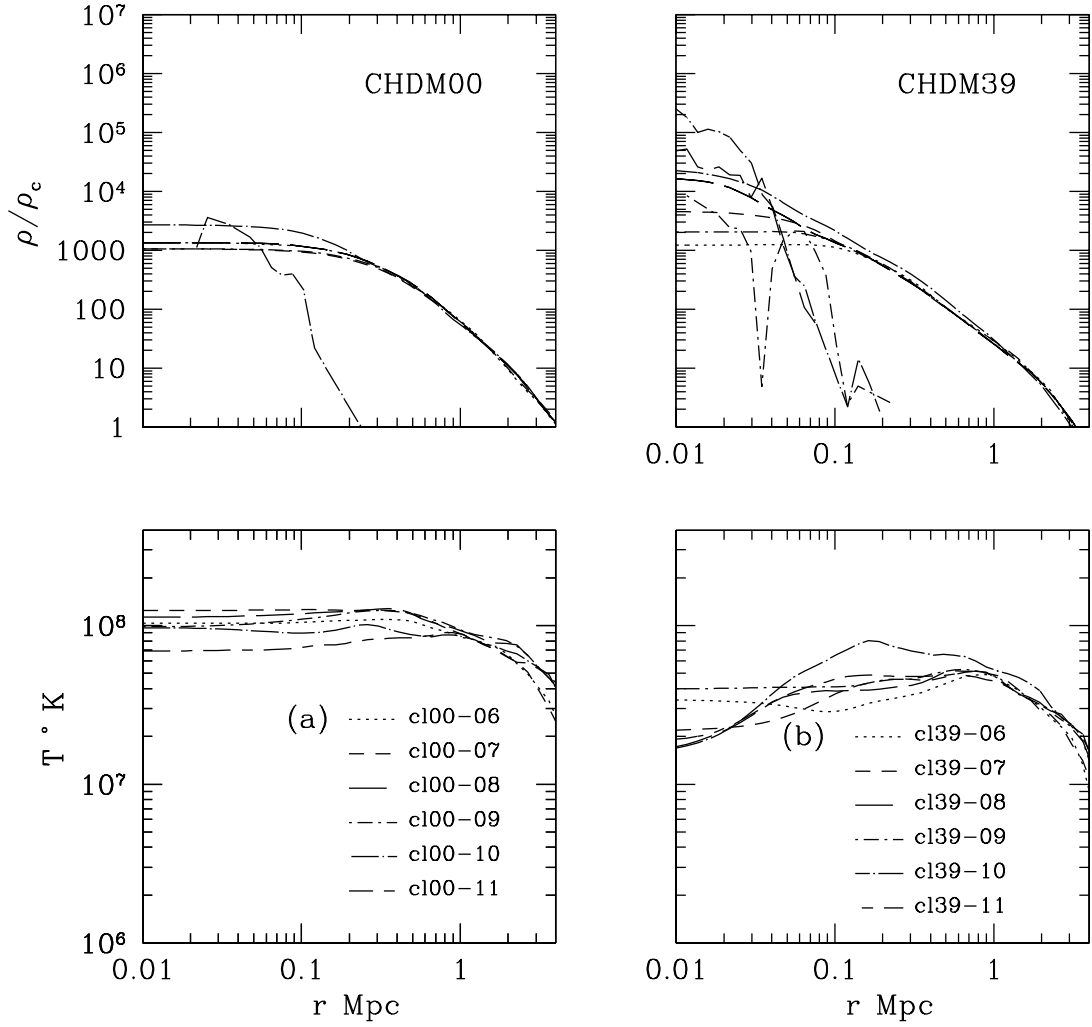} 
 \end{center}
\figcaption[Fig4.eps]{
As in Fig.~\ref{fig:l2a}, but for the cluster simulations shown in 
Fig,~\ref{fig:m1a}.  \label{fig:m2a}}
\end{figure}

\clearpage
\begin{figure}[tbph]
 \begin{center}
  \includegraphics[width=15cm,keepaspectratio]{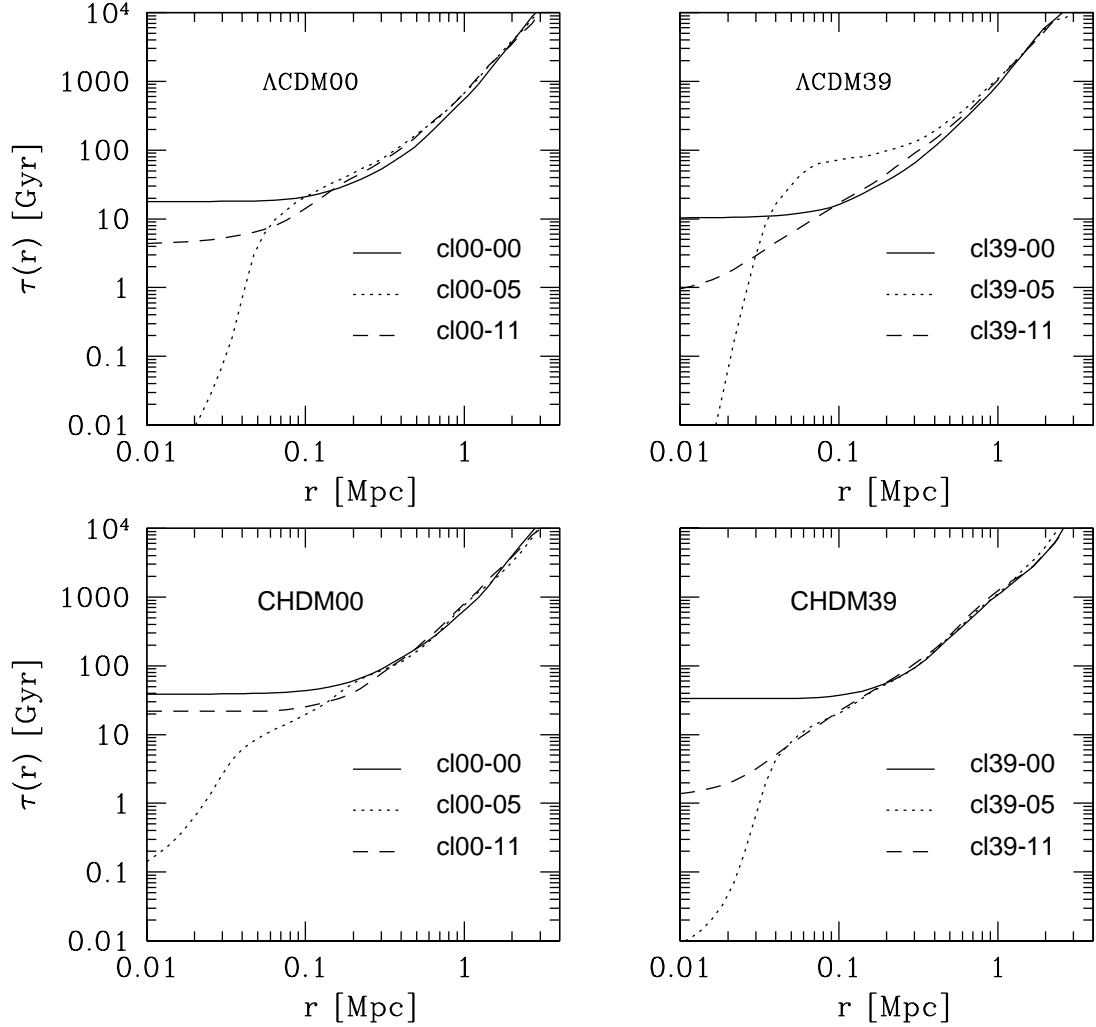} 
 \end{center}
\figcaption[Fig5.eps]{
The cooling time $\tau_c$ as a function of radial distance for the 
four test clusters. $\tau_c$ is defined according to equation \ref{eq:tc}.
In each panel, $\tau_c$ is plotted for three different tests;
the continuous line is the case with no cooling, the dotted line is 
the pure cooling test simulation with the highest resolution 
($-05$), the short-dashed line is  the equivalent simulation run 
but with gas particles being allowed to undergo star formation. \label{fig:tc}}
\end{figure}

\clearpage
\begin{figure}[tbph]
 \begin{center}
  \includegraphics[width=15cm,keepaspectratio]{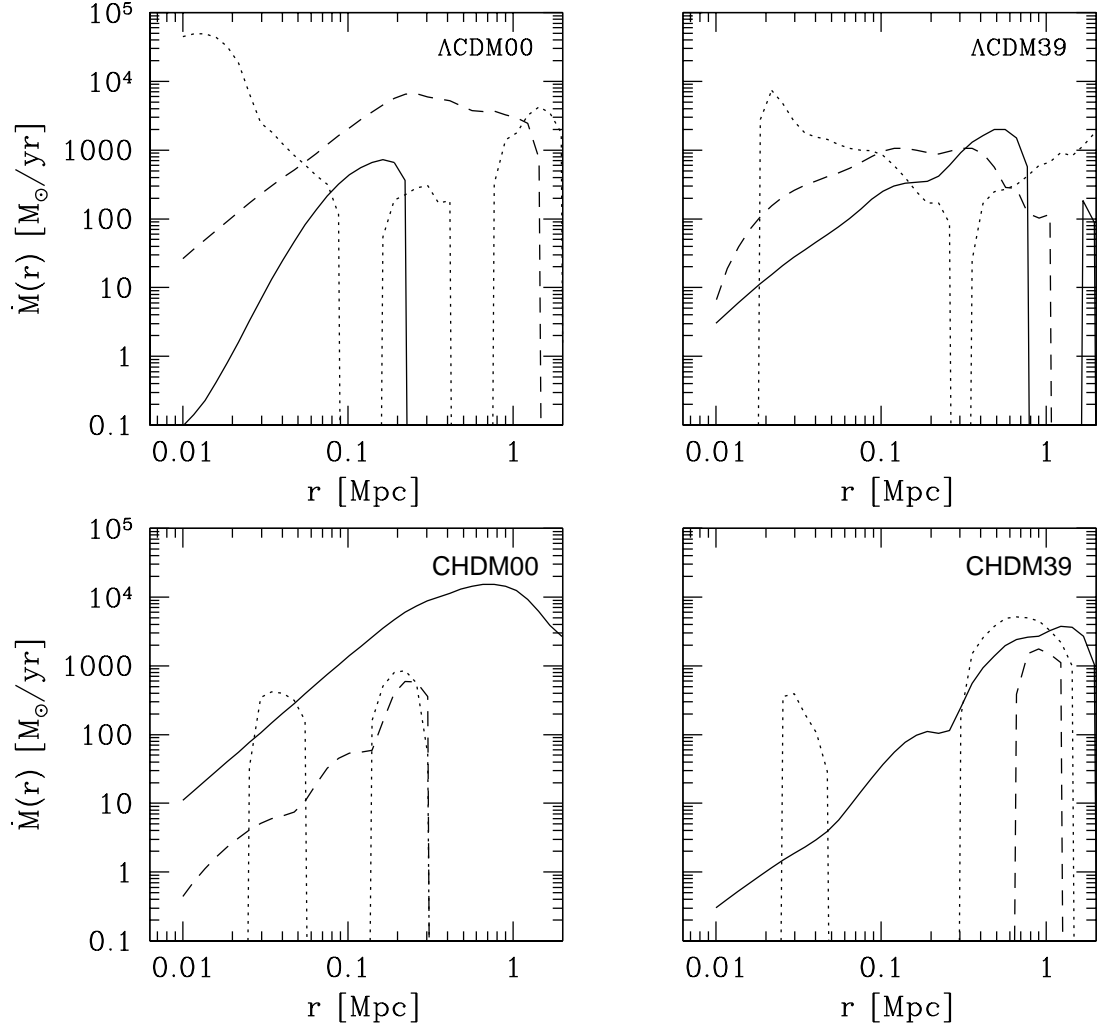} 
 \end{center}
\figcaption[Fig6.eps]{
Mass accretion rates in the four test clusters for the 
same simulation tests shown in Fig.~\ref{fig:tc}. In each radial 
bin, spherical averages for $\dot{M(r)} =4 \pi \rho_g r^2 v_r$ have been plotted
only for negative values of the radial infall velocity $v_r$. \label{fig:md}}
\end{figure}

\clearpage
\begin{figure}[tbph]
 \begin{center}
  \includegraphics[width=15cm,keepaspectratio]{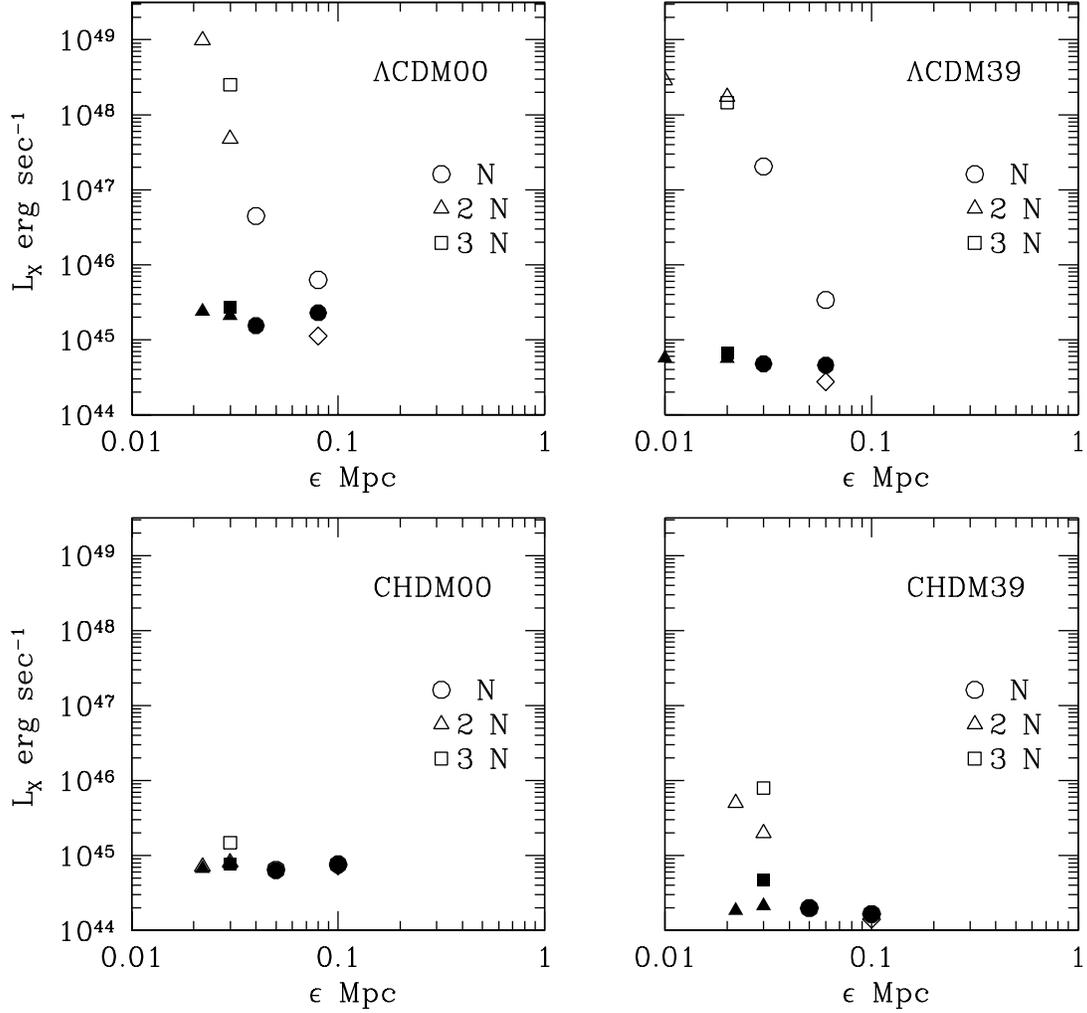} 
 \end{center}
\figcaption[Fig7.eps]{
Final cluster X-ray luminosities as a function of $\epg$ for the
simulation runs performed for the four test clusters. In each panel 
$L_X$ is shown for the 
simulations with gas cooling (open symbols) and for those also including 
 star formation (filled symbols). For these test runs, values 
of the numerical inputs are reported in Tables 2, 3, 4 \& 5.
Simulation clusters with index cl$00-j$ or cl$39-j$ correspond to
 the following symbols: diamond $j=00$; circle $j=01,02$; 
triangle $j=03,04$;  square $j=05$. The filled symbols are the simulations
with index $j+5$. \label{fig:lx}}
\end{figure}

\clearpage
\begin{figure}[tbph]
 \begin{center}
  \includegraphics[width=15cm,keepaspectratio]{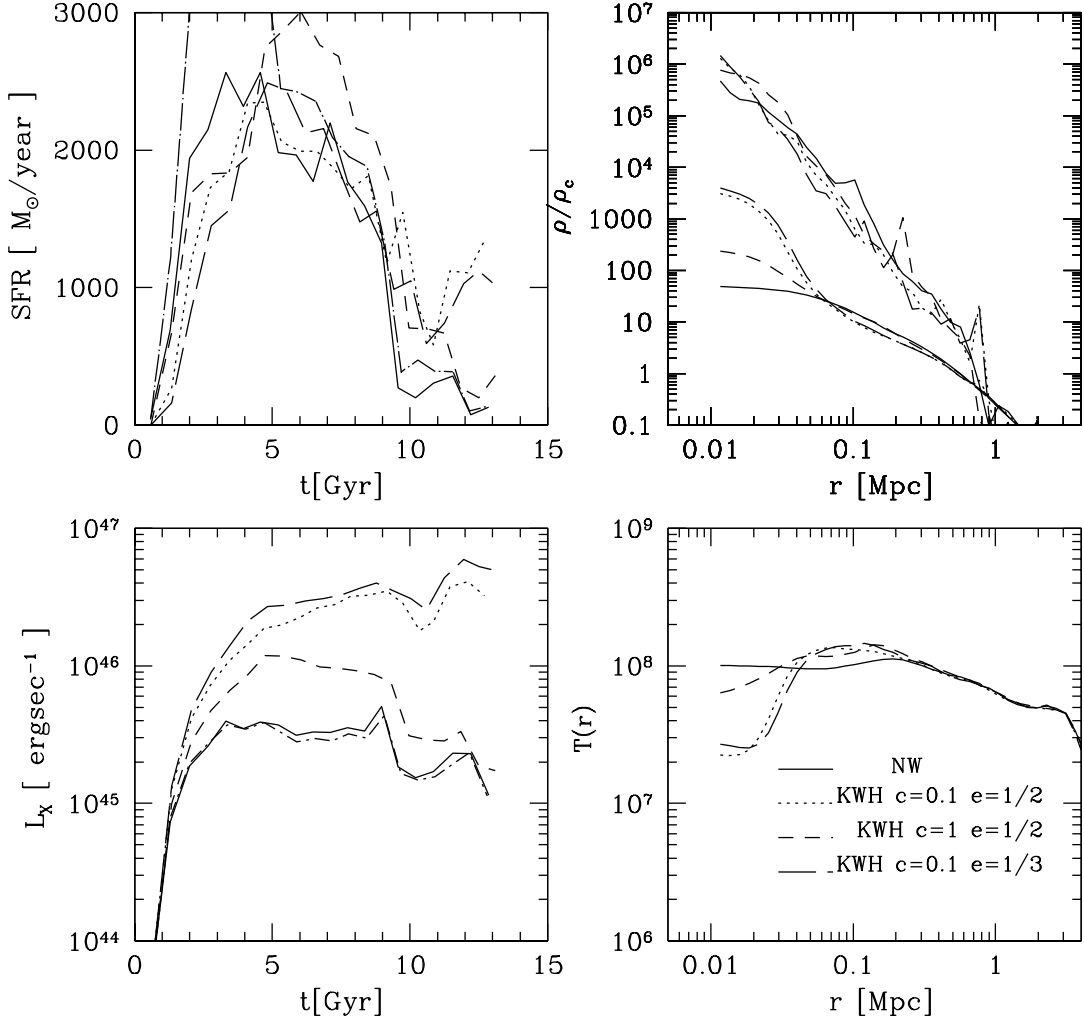} 
 \end{center}
\figcaption[Fig8.eps]{
Plots showing several cluster properties in simulation runs for
the same cluster, but with different SF prescriptions. The test
cluster is \La$00$, and the simulation parameters are those of 
cl$00-11$ (see Table ~\ref{tab:ntestb}).
The different curves are for different SF methods or parameters, as reported
in Table ~\ref{tab:ts1}.
The continuous line refers to the NW method, the others to KWH with 
different $c_{\star}$ and $\eps$ ( $c$ and $e$ in the bottom right panel).
{\it Top left}: star formation rate as a function of time. {\it Bottom left}:
X-ray luminosity versus time. {\it Top right}: final radial density behavior 
for the gas and star components; for the sake of clarity the gas 
densities have been shifted downwards by a factor of 100.
 {\it Bottom right}: Radial temperature profiles.
The dot-dashed line in the two 
left panels is for a simulation run with the NW method, but with an SN 
explosion energy of $\varepsilon_{SN} = 10^{50} erg$. \label{fig:sf}}
\end{figure}

\clearpage
\begin{figure}[tbph]
 \begin{center}
  \includegraphics[width=15cm,keepaspectratio]{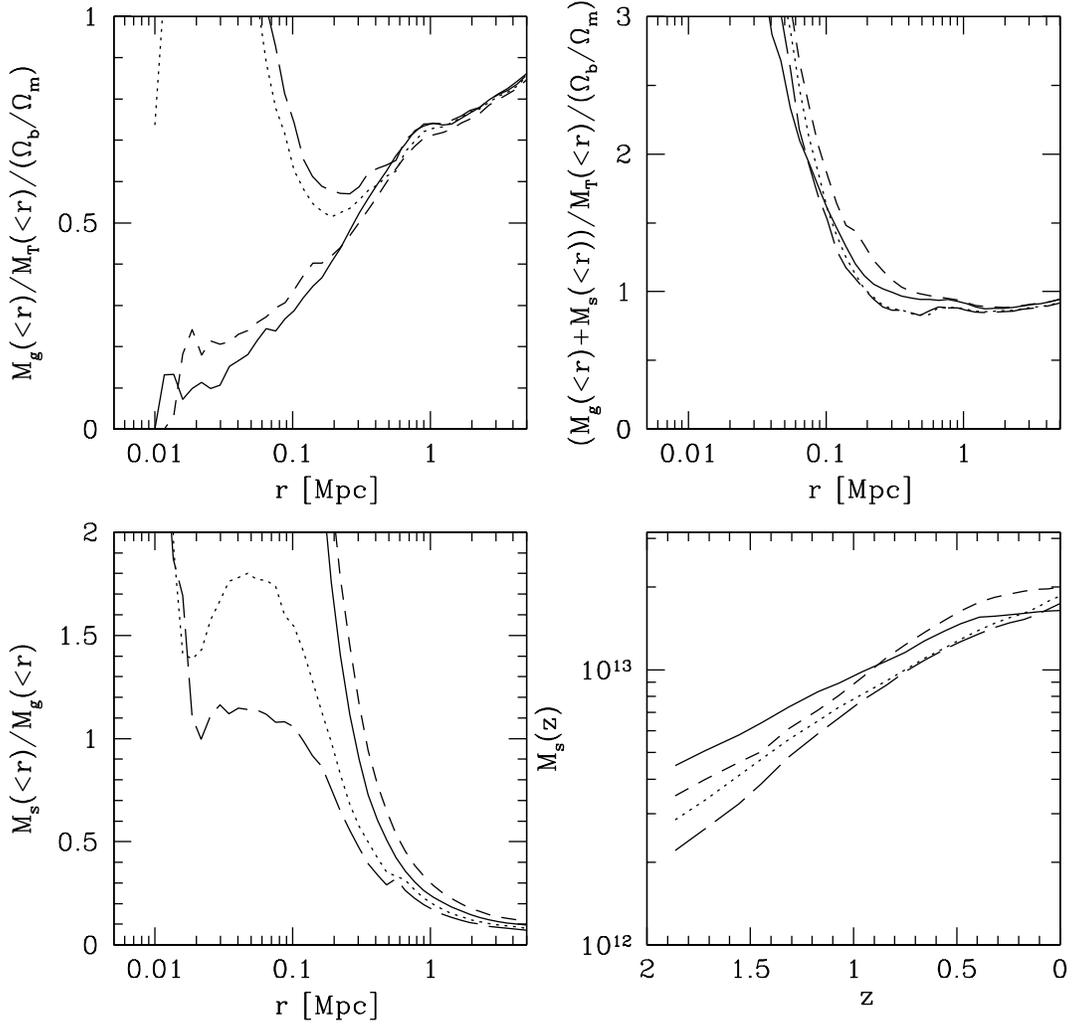} 
 \end{center}
\figcaption[Fig9.eps]{
Baryonic fractions  versus radius for the same simulations
shown in Fig.~\ref{fig:sf}.  
{\it Top left}: ratio of the total mass of gas within the radius $r$, 
to the cumulative cluster mass in units of the universal baryonic fraction.
{\it Top right}: as in the left panel but for the ratio of (star+gas) mass to
the total cluster mass.
{\it Bottom left}: ratio of the star mass within radius $r$ to the 
 mass of gas within $r$.
{\it Bottom right}: total mass produced in stars as a function of redshift.
\label{fig:mg}}
\end{figure}

\clearpage
\begin{figure}[tbph]
 \begin{center}
  \includegraphics[width=15cm,keepaspectratio]{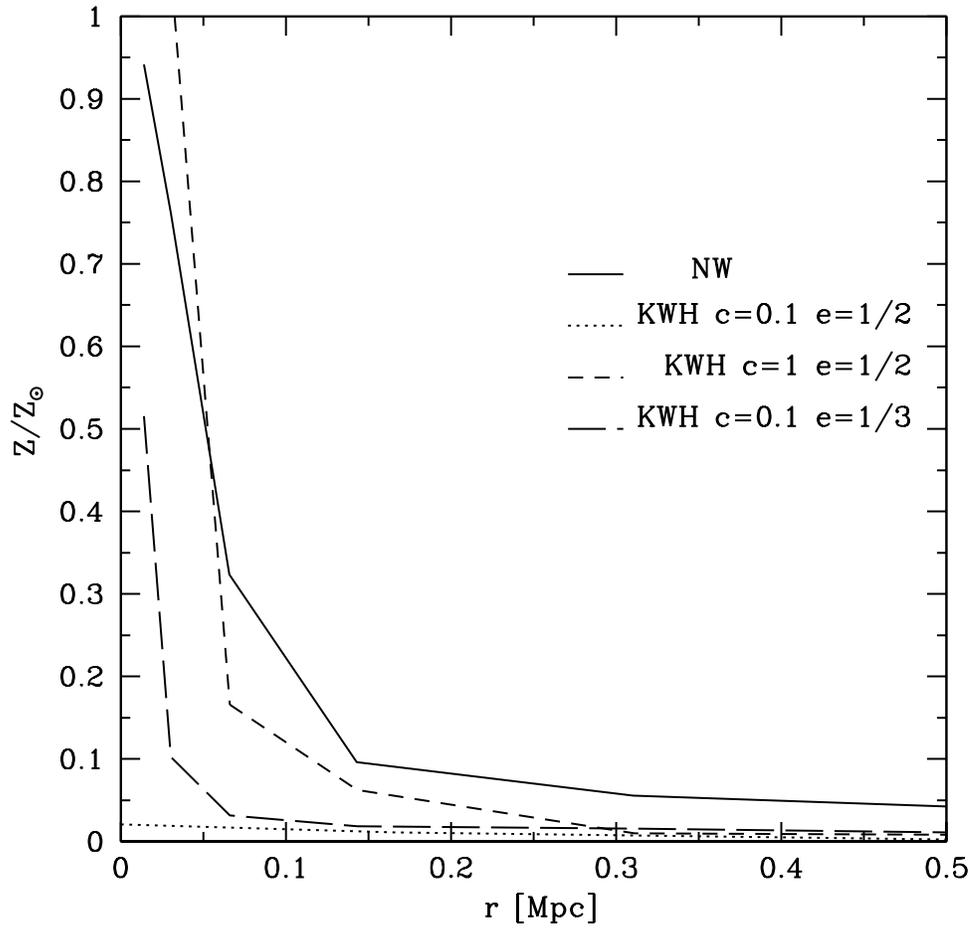} 
 \end{center}
\figcaption[Fig10.eps]{
Spherically averaged iron abundances {$Z=Fe/H$} as a function of radius 
at the final epoch in units of the solar value $4.68\cdot 10^{-5}$.  
The different lines  correspond to the simulations of Fig.~\ref{fig:sf}.
\label{fig:met}}
\end{figure}

\clearpage
\begin{figure}[tbph]
 \begin{center}
  \includegraphics[width=15cm,keepaspectratio]{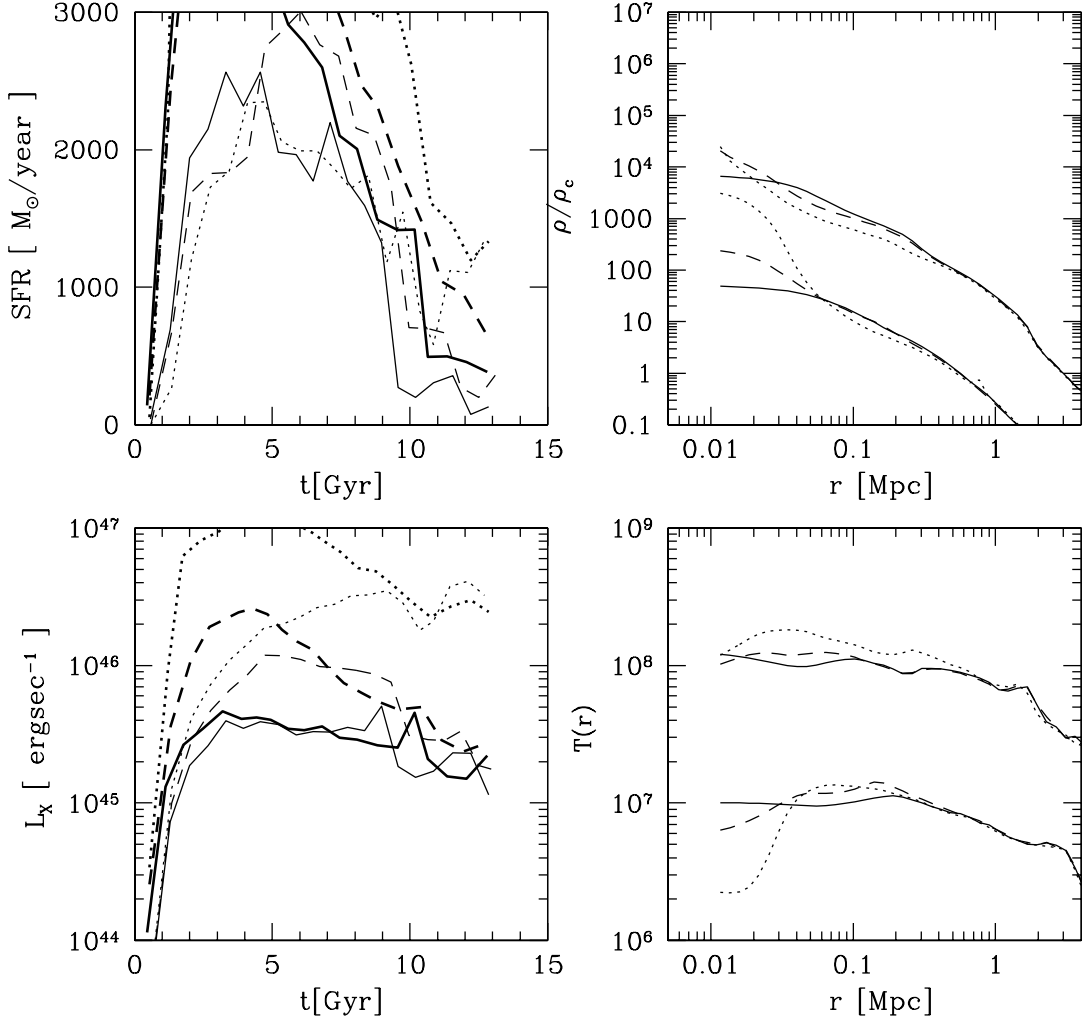} 
 \end{center}
\figcaption[Fig11.eps]{
For the first three runs I, II, III of Table ~\ref{tab:ts1}, the simulation 
results of Fig.~\ref{fig:sf} are compared with the 
corresponding high-resolution runs IH, IIH and IIIH.  
The H simulations  have the same SF parameters as the parent 
simulations, but the numerical parameters are given in the 
last row of Table ~\ref{tab:ntestb}. {\it Left panels }: the thick lines 
correspond to the high resolution simulations.
{\it Right panels}: to facilitate a comparison with the high resolution
results, the radial profiles of Fig.~\ref{fig:sf} have been shifted
downwards by $10^k$, $k=2$ for densities and $k=1$ for temperatures.
\label{fig:sft}}
\end{figure}

\clearpage
\begin{deluxetable}{cccc}
\tabletypesize{\scriptsize}
\tablecaption{Properties of the simulated clusters \label{tab:ta1}}
\tablewidth{0pt}
\tablehead{
\colhead{cluster}&  \colhead{$M_{200}$}& \colhead{$r_{200}$} & 
\colhead{$\sigma_1$}  
}
\startdata
 CHDM00 &   $1.4\cdot 10^{15}$ & 1.83 & 1500  \\
 CHDM39 &  $4.4\cdot 10^{14}$ & 1.25  & 1000 \\
 $\Lambda$CDM00 &   $1.2\cdot 10^{15}$ & 1.98 & 1200  \\
 $\Lambda$CDM39 &   $4\cdot 10^{14}$ & 1.37 & 800  \\
\enddata
\tablecomments{Reference values at $z=0$ for the four 
clusters used in the numerical tests. 
$M_{200}$: cluster mass within 
$r_{200}$ in units of $h^{-1} M_{\odot}$, $r_{200}$ is in units of $h^{-1}$ Mpc,
$\sigma_1$ is the central 1-D dark matter velocity dispersion in 
$Km sec ^{-1}$.
$M_{200}$ is defined as $M_{200}= (4 \pi/3) \Omega_m \rho_c \Delta_c r_{200}^3$,
with $\Delta_c =187 \Omega_m^{-0.55}$ for a flat cosmology. }

\end{deluxetable}

\clearpage
\begin{deluxetable}{cccccccccc}
\tablecolumns{10}  
\tabletypesize{\scriptsize}
\tablecaption{Numerical parameters of the simulations with cooling for 
the \La$00$ cluster \label{tab:ntesta}}
\rotate
\tablewidth{350pt} 
\tablehead{ \colhead{$\Lambda$CDM00} & \colhead{ $\varepsilon_g^{~a}$} &
\colhead{ $m_g^{~b}$} & \colhead{$m_d^{~c}$} & \colhead{$N_g^{~d}$} & 
\colhead{$N_d^{~e}$} &\colhead{ $N_T^{~f}$} & \colhead{$\theta^g$} &
\colhead{ $Q^h$} & \colhead{$z_{in}^{~i}$} } 
\startdata
cl00-00 & 56 & $3.01\cdot10^{10}$ & $2.64\cdot10^{11}$ & 5503 &6295&16463&
0.7& $F$& 10.\\ 
cl00-01 & 56 & $3.01\cdot10^{10}$ & $2.64\cdot10^{11}$ & 5503 &6295&16463&
0.7& $F$ &10.\\ 
cl00-02 & 28 & $3.01\cdot10^{10}$ & $2.64\cdot10^{11}$ & 5503 &6295&16463&
0.7& $F$ &10.\\ 
cl00-03 & 21 &$1.45\cdot10^{10}$ & $1.28\cdot10^{11}$ & 11480&14208&35408&
0.7& $F$& 10.\\ 
cl00-04 &15.4&$1.45\cdot10^{10}$ & $1.28\cdot10^{11}$ & 11480&14208&35408&
0.7& $F$& 10.\\ 
cl00-05&21&$7.47\cdot10^{9}$ & $6.57\cdot10^{10}$ & 22575&25391&67388&
0.7& $F$ &19.\\ 
\enddata
\tablecomments{Simulation parameters of the five test runs for
the $\Lambda$CDM00 cluster. cl$00-00$ is the reference case with no
cooling, taken from VGB. $^{a}$: gravitational softening parameter for the 
gas in $h^{-1}$~Kpc.  $^{b}$: mass of the gas particles in $h^{-1} M_{\odot}$ 
(the cosmology is for $\Omega_m=0.3$ and $h=0.7$).
$^c$ : mass of the dark particles. $^d$: number of gas particles inside
the $L_c/2$ sphere at $z=z_{in}$. $^e$ : as in the previous column but for 
dark particles. $^f$: total number of simulation particles, including 
those in the external shell of radius $L_c$. 
$^g$: value of the treecode gravitational tolerance parameter. 
$^h$ : gravitational quadrupole corrections $F=$ disabled, $T=$ enabled.  
$^i$ : initial redshift for 
the simulation. 
}  
\end{deluxetable}

\clearpage
\begin{deluxetable}{cccccccccc}
\tablecolumns{10}  
\tabletypesize{\scriptsize}
\tablecaption{Numerical parameters for the simulations with cooling and
star formation for the \La$00$ cluster \label{tab:ntestb}}
\rotate
\tablewidth{350pt} 
\tablehead{ \colhead{$\Lambda$CDM00} & \colhead{ $\varepsilon_g$} &
\colhead{ $m_g$} & \colhead{$m_d$} & \colhead{$N_g$} & 
\colhead{$N_d$} &\colhead{ $N_T$} & \colhead{$\theta$} &
\colhead{ $Q$} & \colhead{$z_{in}$} } 
\startdata
cl00-06 & 56 & $3.01\cdot10^{10}$ & $2.64\cdot10^{11}$ & 5503 &6295&16463&
0.7& $F$ &10.\\ 
cl00-07 & 28 & $3.01\cdot10^{10}$ & $2.64\cdot10^{11}$ & 5503 &6295&16463&
0.7& $F$ &10.\\ 
cl00-08 & 21 &$1.45\cdot10^{10}$ & $1.28\cdot10^{11}$ & 11480&14208&35408&
0.7& $F$& 10.\\ 
cl00-09 &15.4&$1.45\cdot10^{10}$ & $1.28\cdot10^{11}$ & 11480&14208&35408&
0.7& $F$& 10.\\ 
cl00-10&21&$7.47\cdot10^{9}$ & $6.57\cdot10^{10}$ & 22575&25391&67388&
0.7& $F$ &19.\\ 
cl00-11&21&$7.47\cdot10^{9}$ & $6.57\cdot10^{10}$ & 22575&25391&67388&
1.0& $T$& 19.\\ 
cl00-11H&10.5&$2.45\cdot10^9$& $2.12\cdot10^{10}$& 69599&74983&204799&$1.0$ 
& $T$ &29.\\
\enddata
\tablecomments{ As in Table~\ref{tab:ntesta}, simulation parameters of the test runs including cooling and star formation for the $\Lambda$CDM00 cluster. 
The index of the run is cl$00-k$, with $k=j+5$ and $j=1,5$ is the 
index of the cooling simulation.  The numerical parameters of the cooling and
 star formation runs are those of the 
corresponding cooling simulation with index $j$.
The last row gives the numerical parameters 
for the high-resolution runs used to test different SF methods. 
}  
\end{deluxetable}

\clearpage
\begin{deluxetable}{cccccccc}
\tabletypesize{\scriptsize}
\tablecaption{  \M$00$ \label{tab:ntestc}}
\tablewidth{0pt}
\tablehead{
\colhead{CHDM00} & \colhead{ $\varepsilon_g$} &\colhead{ $m_g$} & 
\colhead{$m_d^{~a}$} &\colhead{ $N_g$} & \colhead{$N_d^{~a}$} &
\colhead{$N_T$} &\colhead{ $z_{in}$}  
}
\startdata
cl00-00 &50 & $2.28\cdot10^{10}$ & $3.57\cdot10^{11}$ & 5551 &13038&27971&4.8\\ 
cl00-01 &50 & $2.28\cdot10^{10}$ & $3.57\cdot10^{11}$ & 5551 &13038&27971&4.8\\ 
cl00-02 &25 &$2.28\cdot10^{10}$ & $3.57\cdot10^{11}$ & 5551 &13038&27971&4.8\\ 
cl00-03 &15 &$1.1\cdot10^{10}$ & $1.73\cdot10^{11}$ & 11507 &29038&59093&4.8\\ 
cl00-04 &11 &$1.1\cdot10^{10}$ & $1.73\cdot10^{11}$ & 11507 &29038&59093&4.8\\ 
cl00-05 &15 &$5.5\cdot10^{9}$ & $8.9\cdot10^{10}$ & 22575 &50726 &112512&9\\ 
\enddata
\tablecomments{ As in Table ~\ref{tab:ntesta}, but for CHDM$00$. $^a$ : $m_d$ and $N_d$ are the total cold+hot values.}
\end{deluxetable}

\clearpage

\begin{deluxetable}{cccccccc}
\tabletypesize{\scriptsize}
\tablecaption{ \La$39$ and \M$39$ \label{tab:ntestd}}
\tablewidth{0pt}

\tablehead{
\colhead{} & \multicolumn{3}{c}{\La$39$}&\colhead{}& \multicolumn{3}{c}{\M$39$}
\\
\cline{2-4} \cline{6-8} \\
\colhead{} &  \colhead{$\varepsilon_g$} &\colhead{ $m_g$} & \colhead{$m_d$}&
\colhead{} & \colhead{$\varepsilon_g$} &\colhead{ $m_g$} & \colhead{$m_d$}  
}
\startdata
cl39-00 & 42 & $1.47\cdot10^{10}$ & $1.3\cdot10^{11}$ &  
 & 50 & $7.5\cdot10^{9}$ & $1.19\cdot10^{11}$ \\ 
cl39-01 & 42 & $1.47\cdot10^{10}$ & $1.3\cdot10^{11}$  & 
 & 50 & $7.5\cdot10^{9}$ & $1.19\cdot10^{11}$ \\ 
cl39-02 & 21 & $1.47\cdot10^{10}$ & $1.3\cdot10^{11}$  &  
 & 25 & $7.5\cdot10^{9}$ & $1.19\cdot10^{11}$ \\ 
cl39-03 & 14 & $7.2\cdot10^{9}$ & $6.37\cdot10^{10}$  &  
 & 15 & $3.7\cdot10^{9}$ & $5.75\cdot10^{10}$ \\ 
cl39-04 & 10.5 & $7.2\cdot10^{9}$ & $6.37\cdot10^{10}$ &  
 & 11 & $3.7\cdot10^{9}$ & $5.75\cdot10^{10}$ \\ 
cl39-05 &14   &  $3.7\cdot10^{9}$ & $3.22\cdot10^{10}$ &  
 & 15 & $1.89\cdot10^{9}$ & $2.96\cdot10^{10}$ \\ 
\enddata
\tablecomments{ Simulation parameters of the test runs for the clusters 
\La$39$ and \M$39$. The number of particles and initial redshifts are the same 
as for the $00$ clusters.}
\end{deluxetable}
\clearpage

\begin{deluxetable}{cccccccc}
\tabletypesize{\scriptsize}
\tablecaption{  Simulation runs with different star formations
parameters \label{tab:ts1}}
\tablewidth{0pt}
\tablehead{
\colhead{method}& \colhead{ $c_{\star}$}& \colhead{$\eps$} &
\colhead{ $\varepsilon_{SN}$} &\colhead{ $T^a_{em}$} & \colhead{$T^b_m$}
& \colhead{$ T^c_{em} ( < 50 Kpc)/T_{em}$} & \colhead{run} 
}
\startdata
 NW &  1 & 1/2 & 1 & 8.3 & 5.66& 0.08 & I  \\         
 KWH & 0.1 & 1/2 & 1 & 1.9 & 5.5 & 0.8 & II  \\       
 KWH & 1 & 1/2 & 1 & 8.96 & 5.77 & 0.2 & III  \\      
 KWH & 0.1 & 1/3 & 1 & 2.05 & 5.42 & 0.87 & IV  \\
 NW &  1 & 1/2 & 0.1 &  7.88 & 5.54 & 0.08 & V  \\
\enddata
\tablecomments{ SF parameters for the simulations used to compare different
models of SF. The test cluster is \La$00$ and the numerical
parameters are those of cl$00-11$ (see Table ~\ref{tab:ntestb}). The different 
SF parameters are defined in \S2, $\varepsilon_{SN}$ is the 
SN explosion energy in units of $10^{51} erg$. $^a$ : emission weighted 
temperature in $KeV$ units at $z=0$. $^b$ : mass-weighted temperature at $z=0$.
$^c$ :relative contribution to the total emission-weighted temperature from the
$r=50Kpc$ cluster inner region.}
\end{deluxetable}
%
\vfill
\end{document}